\documentclass[aps,prl,twocolumn,superscriptaddress,showpacs,amsmath,nofootinbib]{revtex4}
\usepackage{graphicx}
\usepackage{bm}
\usepackage{color}
\usepackage{enumitem}
\usepackage{amsmath}

\def\lsim{\mathrel{\rlap{\lower4pt\hbox{\hskip1pt$\sim$}}
    \raise1pt\hbox{$<$}}}         
\def\gsim{\mathrel{\rlap{\lower4pt\hbox{\hskip1pt$\sim$}}
    \raise1pt\hbox{$>$}}}         

\begin{document}


\title{Composite Fermions and the First-Landau-Level Fine Structure \\
of the Fractional Quantum Hall Effect}



\author{W. C. Haxton}
\email{haxton@berkeley.edu}
\affiliation{Department of Physics, University of California, and \\
Nuclear Science Division, Lawrence Berkeley National Laboratory,
Berkeley, CA 94720}
\author{Daniel J. Haxton}
\email{djhaxton@lbl.gov}
\affiliation{Chemical Sciences Division, Lawrence Berkeley National Laboratory}


\date{\today}

\begin{abstract}
A set of scalar operators, originally introduced in connection with an analytic first-Landau-level (FLL)
construction of fractional quantum Hall (FQHE) wave functions for the sphere, are employed in a somewhat different way to
generate explicit representations of both hierarchy states (e.g., the series
of fillings $\nu$=1/3, 2/5, 3/7, ... ) and their conjugates ($\nu$ = 1, 2/3, 3/5, ...) 
as non-interacting quasi-electrons filling fine-structure
sub-shells within the FLL.  This yields, for planar and spherical geometries, 
a quasi-electron representation of the incompressible FLL state of filling $p/(2p +1)$ in a
magnetic field of strength B that is algebraically identical to
the IQHE state of filling  $\nu=p$ in a magnetic field of strength $\mathrm{B}/(2p+1)$.   The construction provides
a precise definition of the quasi-electron/composite fermion that differs in some respects from common
descriptions: they are eigenstates of $L,L_z$;  they and the FLL subshells they occupy carry a third index $\mathcal{I}$ that is associated with 
breaking of scalar pairs;
they absorb in their internal wave functions one, not two, units of magnetic flux; and they share a common,
simple structure as vector products of a spinor creating an 
electron and one creating magnetic flux.  We argue that these properties are a consequence 
of the breaking of the degeneracy of noninteracting electrons within the FLL
by the scale-invariant Coulomb potential.
We discuss the sense in which the wave function construction supports
basic ideas of both composite fermion
and hierarchical descriptions of the FQHE.  We describe
symmetries of the quasi-electrons in the $\nu=1/2$ limit, where a deep Fermi sea of quasi-electrons forms, and
the quasi-electrons take on Majorana and pseudo-Dirac characters.  Finally, we show that the wave functions
can be viewed as fermionic excitations of the bosonic half-filled shell, producing at $\nu=1/2$ an operator
that differs from but plays the same role as the Pfaffian.
  \end{abstract}


\maketitle


\section{I.  Introduction}
Twenty years after the discovery \cite{Tsui} of the fractional quantum Hall effect (FQHE) Dyakonov \cite{DY} wrote a 
rather sharp critique of the present state of its theory.  Among his criticisms were the
absence of any simple or beautiful first-Landau-level (FLL) formulation of the hierarchy
wave functions ($\nu<1/2$), the lack of an explicit representation of conjugate states (e.g., fillings where $\nu >1/2$, so $\nu$ =2/3, 3/5, etc.),
the absence of a justification for wave functions in terms of underlying principle
such as minimizing the interaction energy, and the lack of a
sound theoretical foundation for concepts such as composite fermions (CFs), which he argued 
had been neither derived nor even adequately defined.   While recognizing
the phenomenological success of Jain's ``wave function engineering" from which the
notion of CFs \cite{CFs} derives, he argued that 
far too many simple questions about the FQHE remain without reasonable answers.

Here we address some of Dyakonov's concerns by providing an explicit mapping from the
strongly-correlated-electron form of FQHE wave functions to 
a quasi-electron (or CF) form, yielding
wave functions algebraically identical to those of the noninteracting
integer QHE.  This is done for both the sphere and plane.
The quasi-electrons have a simple analytic form and are eigenstates of
the angular momentum operators $L$ and $L_z$, and also carry a label $\mathcal{I}$
related to Haldane's $p$-wave pseudopotential \cite{Haldane}.   The quasi-electrons
occupy fine-structure FLL sub-shells distinguished by the same label.  The wave functions
for $\nu=1/3, 2/5, 3/7, ... (p/2p+1)$ and  $\nu = 1, 2/3, 3/5, ... (p/(2p-1)$  correspond to configurations where
$p$ sub-shells are fully filled by their respective quasi-electrons.  The sub-shell
structure is induced by the Coulomb interaction, with the gap between neighboring 
sub-shells reflecting the energy cost of removing one
$p$-wave coupling between quasi-electron 1 and one of its $N-1$ neighbors.
We discuss the relevance of the construction to a number of current
questions about the FQHE, including the relationship between CF 
and hierarchical descriptions of wave functions \cite{jainhier}; the structure of the
$\nu=1/2$ state  \cite{Son,HLR,MR}, where we argue there exist alternative Majorana and
pseudo-Dirac descriptions associated with the symmetries at this filling, and where we identify
an operator quite similar to a Pfaffian; and possible
systematic improvements of wave functions, in the spirit of an effective theory, that may
help address certain open questions about the FQHE.

Jain constructed his 2/5, 3/7, 4/9, ... hierarchy wave functions by first operating with a multiply-filled integer quantum Hall
state on the half-filled symmetric state, producing a wave function spread
over multiple higher Landau levels, which then was projected numerically onto
the FLL, eliminating the unwanted components.   When tested against numerical
solutions obtained by diagonalizing the Coulomb interaction, excellent overlaps were found,
comparable to those for Laughlin's \cite{Laughlin} $\nu$= 1/3 and 1/5 wave functions. 

However, as Dyakonov discusses, the Jain construction is troubling on several grounds.  
Laughlin's $\nu$= 1/3 and 1/5 states are supported by certain variational arguments.  For example,
at short-distances the wave functions vary as $(z_i-z_j)^3$ and $(z_i-z_j)^5$, eliminating the
most repulsive multipoles of the electron-electron Coulomb interaction $\sim 1/|z_i-z_j|$.
Jain's construction makes no reference to the electron-electron interaction, but instead employs an operator
taken from the noninteracting integer quantum Hall effect (IQHE), with electrons
occupying higher LLs characterized by large magnetic gaps,
which at the end are eliminated numerically  --
a procedure Dyakonov termed ``bizarre."   Laughlin's wave function can be
expressed analytically as simple products of closed-shell operators, while Jain's final result has
no such representation.  

Laughlin's construction is based on the rescaling of all inter-electron correlations,
for $\nu=1,1/3, 2/5$,  by factors of $(z_i-z_j)^m$, 
$m=1,3,5$,  a procedure that reflects the scale invariance of the underlying
Coulomb potential.  In 1996 Ginocchio and Haxton \cite{GH} (GH) generalized this approach to successively
larger groups of electrons: recognizing that higher density ``defects" would necessarily arise beyond $\nu=1/3$ --
beyond this filling $p$-wave electron pairs must start to appear --
they introduced operators to create such defects, then looked for solutions that would distribute these over-dense regions
uniformly.
On the sphere the closed-shell operators Laughlin employed can be labeled by the quantum number $\ell$,
where $2\ell+1=N$ is the electron number.  The GH construction produces a 
larger class of such scalar operators, defined by two quantum numbers $\ell$ and $s$.  The GH and Laughlin
operators coincide for $s$=0.  The GH operators generate the full set of hierarchy states (the Jain states)
when $s=0,1/2,1,....$ is allowed to run, constrained by
 $s \le \ell$; and they generate hierarchy states and their conjugates when $l,s$ are varied, without constraints.  
 These quantum numbers are related to the electron number by
$N=(2s+1)(2l+1)$.  

The GH operators were later used by Jain and Kamilla \cite{JK}, but otherwise have not
been broadly applied.  One reason may be the limitation to the sphere: GH used this geometry
because spherical $N$-electron wave functions generated from scalar operators have total angular momentum $L=0$,
guaranteeing both translational invariance and homogeneity (uniform one-body density).
But most investigators work
in the plane, where simple polynomials replace angular momentum couplings. 
Second, while the GH treatment of higher-order electron correlations led to operators with a manifest
sub-shell structure, implying a fine-structure splitting of the FLL, this structure was not
obvious in the final wave functions: The GH operators contain
spherical tensor derivatives (operators that destroy magnetic flux) that
can be evaluated analytically, but a compact form of
the scalar wave function with all such derivates eliminated by such evaluation was not provided.  This
shortcoming complicates the construction of analogous wave functions in the plane,
as derivatives on the sphere are associated with raising and lowering operators 
that operate only within finite Hilbert spaces, unlike the case of the plane.
Although the GH construction addresses two issues Dyakonov raised --  producing an
analytic generalization of Laughlin's construction with the same variation justification,
and generating the full sets of hierarchy and conjugate states -- the final
wave functions lack the elegant simplicity of Laughlin's results.

GH, in fact, did not seek the most simple application of their operators, as their goal
was to account for Jain's surprising construction.  Thus they applied their $\ell s$
operators as Jain did his IQHE operators, acting directly on the half-filled shell.  This procedure
yields results numerically identical to Jain's, and demonstrates that correlations within
the FLL generate an SU(2) sub-shell algebra distinct from but algebraic identical
to that of multiply filled IQHE states.  This algebraic similarity accounts for Jain's success,
though his construction misses the connection between broken scalar pairs and the
generation of angular momentum that is responsible for the GH FLL sub-shell structure.

In this paper we discuss an alternative\footnote[1]{We denote
wave functions obtained with the original use of the GH operators as GH wave functions, and those
with the current formulation the GH$^2$ wave functions.} use of GH operators, as
creators of
quasi-electrons.\footnote[2]{We use the term quasi-electron, rather than ``composite fermion," as
the latter is most commonly described as an electron
coupled to two (or an even number of) units of magnetic flux.   In our treatment, such objects arise in recursion operators:
the ($N$+1)-electron $\nu=1/3$ state can be generated from the $N$-electron state by a recursion
operator identical to that for the $\nu=1$ state, except that the electron in the latter is replaced by
a composite object, a single-particle state coupled to two units of magnetic flux. But these are not
the objects that form the single-Slater-determinant representation of the hierarchy states:  these are
electrons coupled to one unit of magnetic flux.}  The resulting GH$^2$ hierarchy and conjugate states are
closed-shell configurations of families of quasi-electrons, all of which 
have a common form as tensor products of two spinors, one creating an electron and one creating a
single unit of magnetic flux. 
 As the construction eliminates all GH derivatives,
wave functions can be readily expressed in either spherical or planar geometry.
The GH$^2$ quasi-electron wave function of filling $\nu=p/(2p+1)$ and magnetic field strength $B$ is
identical in form to the
IQHE electron wave function of the same electron number, but in a reduced magnetic field $B/(2p+1)$.

The plan of this paper is as follows.  In Sec. II we review properties of Laughlin's wave functions, 
providing benchmarks for subsequent discussions of other
hierarchy states.  We use the Laughlin wave function to illustrate how planar operators can be written
as analogs of spherical ones, and use this mapping to define states of uniform density in
the plane -- otherwise an ill-defined concept, in our view.  This involves reorganizing the
usual planar degrees of freedom into single-particle and Schur-polynomial spinors, with the latter
representing the addition of a unit of magnetic flux.  We describe how vector products can be formed in
the plane to produce quasi-electrons with good angular momentum, and consequently how to generate
planar ``scalars" that are both translationally invariant and uniform in density.

In Sec. III we discuss the GH operator construction in more detail than was possible
in the original letter \cite{GH}.   We describe the $\ell s$ form, connected with electron correlations,
and the $(\ell s)j$ form, connected with FLL shell fine structure. The two 
representations allow one to understand the connections between
FLL electron correlations, energy minimization, and sub-shell structure.

In Section IV we show how the
GH operators can be used to produce hierarchy and conjugate states in their GH$^2$ form,
noninteracting quasi-electrons occupying filled sub-shells.  We describe properties
of the quasi-electrons, then properties of the low-momentum representation of the many-electron
states that can be built as closed-shell configurations of the quasi-electrons.  We illustrate the novel
properties of the sub-shell structure, which is not static but
evolves as electrons are added, considering various ``trajectories"
in the two-parameter Hilbert space, such as fixed $\nu$ with increasing $N$, or fixed magnetic field
strength with increasing $N$.  We describe symmetries associated with in the exchange of the
particle and flux spinors for the  $\nu=1/2$ case, where the quasi-electrons exhibit special
symmetries: we find Majorana-like and Dirac-like
solutions at $\nu=1/2$, the latter associated with spin flip. We also show the connection to
the Pfaffian.  We evaluate the overlaps of GH$^2$ wave functions
with results from exact diagonalizations
of the Coulomb interaction.  
The concluding Section V includes a discussion of issues for further study.   The formalism is 
suggestive of an effective theory, and we remark
on work that could be undertaken to explore this possibility, including potential
connections to states at fillings like $\nu=4/11$.

In the Appendix, we present a more technical discussion of correlations, to contrast the current
construction (which is guided by the scale invariance of the Coulomb potential) with alternative
variational schemes focused of the short wave behavior of wave functions, e.g., some 
generalization of the Haldane pseudopotential for interactions among multiple electrons.

\section{II.  Laughlin's Wave Function}
\noindent
{\bf Single electron states in the plane:} The Hamiltonian for an electron moving in a plane under the influence of a
perpendicular magnetic field $B \hat{z}$ is
\begin{equation}
\label{eq:Ham0}
H =  {1 \over 2m}  \left| {\hbar \over i} \vec{\nabla} - {q \over c} \vec{A} \right|^2
\end{equation}
where $m$ is the electron mass and $q=-e=-|e|$ is the electron charge.   We take the direction of electron rotation to be clockwise in the $\hat{x}-\hat{y}$ plane
(as viewed in a right-handed coordinate system from positive $\hat{z}$), a choice that 
requires B to be negative -- the field points in the $-\hat{z}$ direction.  Thus
$qB=e|B|$ is positive.  In the symmetric 
gauge we employ $\vec{A} = B (x \hat{y}-y \hat{x})/2$.  

The following operators can be defined in terms of the dimensionless
coordinate $z=(x+ i y)/a_0\sqrt{2} =r e^{i \theta}$, where the 
magnetic length $a_0 = \sqrt{\hbar c /e |B|}$, and  its conjugate $z^* =(x- i y)/a_0\sqrt{2}$,
\begin{eqnarray}
\label{eq:sq}
a&=& -i \left({\partial \over \partial z^*} + {z \over 2} \right)~~~~a^\dagger = i \left( -{\partial \over \partial z} + {z^* \over 2} \right) \nonumber \\
b &=&  {\partial \over \partial z} + {z^* \over 2} ~~~~~~~~~~~~~
b^\dagger = - {\partial \over \partial z^*} + {z \over 2} 
\end{eqnarray}
The one-body part of Eq. (\ref{eq:Ham}) can then be written
\begin{equation}
H = \hbar \omega \left( a^\dagger a +{1 \over 2} \right) 
\end{equation}
 where the cyclotron frequency $\omega_c = e|B|/ m c$.
The  normalized and degenerate single-electron states of the FLL
with energy $\hbar \omega_c/2$ are
\begin{equation}
\langle z | k \rangle={1 \over \sqrt{ \pi k!}} z^k e^{-|z|^2/2} = {1 \over \sqrt{ \pi k!}} r^k e^{i k \theta} e^{-r^2/2} 
\label{eq:psp}
\end{equation}
These states can be generated as follows
\begin{equation}
\label{eq:spstate}
| k \rangle = { 1 \over \sqrt{k!}} (b^\dagger)^k |k=0 \rangle~~~\langle z | k=0 \rangle = {1 \over \sqrt{ \pi}} e^{-|z|^2/2}.
\end{equation}
yielding the raising and lowering relations
\begin{equation}
b^\dagger |k \rangle = \sqrt{k+1} | k+1 \rangle~~~~~~~b | k\rangle = \sqrt{k} |k-1 \rangle
\end{equation}
as well as
\begin{equation}
  b |k=0 \rangle = 0~~~~~[b,b^\dagger]=1.
  \end{equation}
The single-particle states are eigenstates of the orbital angular momentum operator $L_z$ with
eigenvalue $k$,
\begin{equation}
L_z = \left[ \vec{r} \times \vec{p} \right]_z=\hbar( b^\dagger b-a^\dagger a) \begin{array}{c} ~~ \\ \longrightarrow \\ {}^{FLL} \end{array}  \hbar~ b^\dagger b ~~~~~
b^\dagger b | k \rangle = k |k \rangle
\end{equation}

\noindent
{\bf Single electron states on the sphere:} 
The original  GH operator construction was done on the sphere, a FQHE geometry
introduced by Haldane \cite{Haldane}:
the electrons move on the sphere's surface under the influence of a radial
magnetic field generated by a Dirac monopole at the origin. 
This geometry provides two advantages.  First,  in contrast to the plane, 
there is both a defined surface area and a fixed number of FLL single-particle states, determined
by the number of monopole quanta.  Thus densities and fractional fillings can be defined
unambiguously.\footnote[3]{  In the plane for finite $N$, however, there is less clarity -- disks have soft edges,
making it difficult to formulate a crisp definition of density without handwaving 
about regions within a magnetic length or two of the edge.  This ambiguity carries 
over to the definition of an appropriate many-body Hilbert space at finite $N$:  one can
envision truncating on the number of quanta $k$ in single-particle states,
or, alternatively, in the total number of quanta in many-body states.  Thus
Haldane remarks on the uniqueness of the $N$=3 Laughlin state on the
sphere, yet ``in planar geometry, Laughlin's $N$=3 droplet states are reportedly not exact."}
Second, on the sphere many-electron states with total $L$=0
are both homogeneous -- uniform density over the sphere -- and translationally invariant 
(displacements generated by $L_x$ and $L_y$) \cite{GHsym}.  States with $L=0$ can
be constructed from single-particle spinors of good $L$, $L_z$ (rotation matrices),
using standard angular momentum coupling methods.

Dirac's monopole quantization condition requires the total magnetic flux through the sphere of radius $R$ to
be an integral multipole of the elementary flux $\Phi_0 = h c/q$,
$\Phi = 2 S \Phi_0$. The Hamiltonian for an electron
confined to the sphere is
\begin{equation}
\label{eq:Ham0sp}
H = {1 \over 2m S a_0^2}  \left| \vec{r} \times \left({\hbar \over i} \vec{\nabla} - {q \over c} \vec{A}  \right)\right|^2 \nonumber \\
= {\omega \over 2S  \hbar} \vec{\Lambda}^2
\end{equation}
where $\vec{\Lambda}=\vec{r} \times \left({\hbar \over i} \vec{\nabla} - {q \over c} \vec{A} \right)$ is the dynamical
angular momentum, $\omega=q B/mc = \hbar/m a_0^2$ is the cyclotron frequency, and
$\vec{\nabla} \times \vec{A} = B \hat{\Omega}$ where $\hat{\Omega} \equiv \vec{r}/R$.  The 
angular momentum operators $\vec{L}=\vec{\Lambda}+ \hbar S \vec{\Omega}$
satisfy the commutation relations $[L_i,L_j]=i \epsilon_{ijk} L_k$.  As $\vec{\Lambda}$ is normal to the
surface while $\vec{\Omega}$ is radial, $\vec{\Omega} \cdot \vec{\Lambda} =\vec{\Lambda} \cdot \vec{\Omega}=0$ 
and $\vec{L} \cdot \vec{\Omega} = \vec{\Omega} \cdot \vec{L} = 0$.  These relations 
give $\vec{\Lambda}^2 = \vec{L}^2 - \hbar^2 S^2=\hbar^2( L(L+1)-S^2)$, where $L=S, S+1, S+2, ...$.  
Consequently the eigenvalues corresponding to the Landau levels are
\[ E = {\hbar \omega \over 2S} (L(L+1)-S^2),~~L=S,S+1,S+2,... \]
The single-particle wave functions are the Wigner D-functions 
\begin{equation}
 \mathcal{D}^L_{S,M}(\phi,\theta,0), ~~-L \le M \le L
 \label{eq:D}
 \end{equation}
Thus there are $2S+1$ degenerate single-particle states in the FLL, $2S+3$ in the second LL, etc.
The FLL wave functions can be written as a monomial of power $2S$ in the elementary spinors $u_m$,
\begin{equation}
\mathcal{D}^S_{S,M} = \left[ {(2S)! \over (S+M)!(S-M)!} \right]^{1/2} u_{1/2}^{S+M} u_{-1/2}^{S-M}\equiv [u]^{S}_M
\label{eq:us}
\end{equation}
where 
\begin{equation}
u_m(\phi,\theta)=\mathcal{D}_{1/2,m}^{1/2} (\phi,\theta,0) 
= \left\{ \begin{array}{ll} \cos{\theta/2}~ e^{i \phi/2}, & m=~~{1 \over 2} \\ ~ & ~ \\ \sin{\theta/2}~e^{-i \phi/2}, & m=-{1 \over 2} \end{array} \right. 
\label{eq:elementary}
\end{equation}\\

\noindent
{\bf Jain's operator in spherical notation:} As we will discuss in Sec. III, Jain used
the antisymmetric IQHE state consisting of the lowest $2s+1$ shells, fully occupied, as an operator, acting on the
half-filled shell.  Thus
$L$ in Eq. (\ref{eq:D}) runs over
$S \le L \le S+2s$, where $s$ is an integer or half integer, $s \ge 0$.  Defining $ l \equiv S+s$, so that
$l$ can also be an integer or half integer,
the single-electron spinors (Eq. (\ref{eq:D})) for these shells are
\begin{equation}
 \mathcal{D}^{(ls)j}_{l-s,m_j}(\phi,\theta,0) ~~~~~ l-s \le j \le l+s,~~-j \le m_j \le j
 \label{eq:DJ}
 \end{equation}
 Note that $l  \ge s$.
There are $(2l+1)(2s+1)$ allowed values of $j,m_j$.\\

\noindent
{\bf The interacting problem:}
The planar Hamiltonian responsible for the FQHE effect is obtained by adding the electron-electron Coulomb interaction
to the N-electron version of Eq. (\ref{eq:Ham0}),  as well as a uniform neutralizing electrostatic background 
field $V_j$
\begin{equation}
\label{eq:Ham}
H = \sum_{j=1}^N \left(  {1 \over 2m}  \left| {\hbar \over i} \vec{\nabla}_j - {q \over c} \vec{A} \right|^2 + V_j) \right) + {1 \over 2} \sum_{i,j=1}^A {q^2 \over |\vec{r}_i-\vec{r}_j|}
\end{equation}
On the sphere, $ |\vec{r}_i-\vec{r}_j|$ can be identified with the chord separation
of electrons $i$ and $j$.  The many-electron Hilbert space consists of Slater determinants form
from these single-particle wave functions defined above.  The degeneracy among these states is the broken
by the interaction.
In the case of the sphere, as the FLL single-particle basis is of dimension $2S+1$, the fraction of the single
particles states filled is $N/(2S+1)$.  The fractional filling $\nu$ of a series of related states, such as the 
Laughlin $m=3$ series discussed below, is defined as the large-$N$ limit of this ratio.\\

\noindent
{\bf Plane-sphere relationships and scalar contractions:}  On the sphere many-body states of definite
total $L=0$ are both rotationally invariant (that is, invariant under small displacements along the
sphere's surface) and homogeneous (uniform one-electron density).  
In order to generalize the GH spherical construction to the plane, it is important to find a procedure for
generating analogous scalar states on the plane.  Such states will then be automatically translationally invariant 
and can also be considered homogeneous, as we discuss below.  This requires
an analog of the spherical tensor product, in which objects of definite rotational
symmetry are combined to produce new objects with such symmetry, 
\[ U_{LM}, V_{L^\prime M^\prime} \rightarrow  \left[ U_L \otimes V_{L^\prime} \right]_{L_0 M_0}. \]

The spherical and planar geometries are related by the mapping of the three-dimensional
rotation group into the two-dimensional Euclidean group
\begin{equation}
\label{eq:group}
L_x \begin{array}{c} ~~ \\ \longrightarrow \\ {}^{R \rightarrow \infty} \end{array}- R P_y~~~~~
L_y \begin{array}{c} ~~ \\ \longrightarrow \\ {}^{R \rightarrow \infty} \end{array}  R P_x~~~~~
L_z \begin{array}{c} ~~ \\ \longrightarrow \\ {}^{R \rightarrow \infty} \end{array} L_z
\end{equation}
The correspondence between $L_x,~L_y$ and $P_x,~P_y$ means that the scalar product we seek
will automatically produce translationally invariant states in the plane.  (That is, the Slater determinants 
one forms will have polynomials that depend only on coordinate differences, while the center-of-mass 
of the $N$-electron system will be in the lowest harmonic oscillator state.  Consequently, while the wave
functions technically involve $2N$ spatial coordinates, in fact they can be considered functions
of just $2(N-1)$ intrinsic coordinates.)  While we have noted that the notion of a homogeneous finite-$N$
state in the plane is ambiguous -- the electrons are not strictly confined to any definite area - the
mapping between sphere and plane can be used to define homogeneity: a planar state is 
homogeneous if it corresponds to a homogeneous spherical state under this mapping.
 
Most discussions of the FQHE on the plane use single-electron wave functions $|k \rangle$, but the
discussion above argues for using objects analogous to the angular momentum spinors of GH, which we
introduce here.   Two kinds of objects are need.  The first, analogous to $\{ \mathcal{D}^S_{S,M} \}$, is the single electron spinor of
 rank $k/2$ with $k+1$ components,\footnote[4]{Despite some potential for confusion we retain the
 conventional definition of $L_z$ so that
 \begin{equation}
 L_z~ [z]^{k /2}_m e^{-|z|^2/2} = \hbar ({k \over 2}+m)~ [z]^{k / 2}_m e^{-|z|^2/2}.\nonumber 
  \end{equation}
 The magnetic quantum number corresponding to $L_z$ is ${k \over 2} + m$, so that it ranges from 
 $k$ to 0 for the components of Eq. (\ref{eq:vector}).}
\begin{eqnarray}
\label{eq:vector}
\left[ z\right]^{k/2}\equiv  \left( \begin{array}{c} {[z]}^{k/2}_{k/2}~~~ \\  {[z]}^{k/2}_{k/2-1} \\ \vdots~~~~ \\ {[z]}^{k/2}_{-k/2}~~~ \end{array} \right)  \equiv  {1 \over \sqrt{k!}}\left( \begin{array}{c} z^{k} \\ k z^{k-1} \\ \vdots \\ k! \end{array} \right) \nonumber \\
\left [z \right]^{k/2}_{m-1} = { d \over dz} ~[z]^{k/2}_m  ~~~~~~~~~~~~~~~~~~
\end{eqnarray}
Effectively we have introduced an angular momentum spinor with its $2(k/2)+1$ magnetic components
as a means of truncating the planar Hilbert space.

One can define a scalar product
between two such planar vectors of the same rank $k$ by
\begin{equation}
\label{eq:scalar}
[z_1]^{k} \odot [z_2]^{k} = \sum_{i=-k}^{k}~(-1)^{k-i}~ [z_1]^{k}_i ~ [z_2]^{k}_{-i} ~.
\end{equation}
It follows simply that this scalar quantity is translationally invariant -- it cannot be lowered,
\begin{equation}
\left( {\partial \over \partial z_1} + {\partial \over \partial z_2} \right) [z_1]^{k} \odot [z_2]^{k}  = 0.
\end{equation}
Simple examples are found in Laughlin's two-electron building blocks of the next subsection
\begin{eqnarray}
&&[z_1]^{1/2} \odot [z_2]^{1/2} =z_1-z_2 \nonumber \\
&&[z_1]^{m/2} \odot [z_2]^{m/2} = ([z_1]^{1/2} \odot [z_2]^{1/2} )^m=(z_1-z_2)^m \nonumber
\label{eq:sc}
\end{eqnarray}

A second kind of spinor of rank $k/2$, symmetric under electron interchange and analogous to the aligned spherical vector 
$[u_1 u_2 \cdots u_k]^{k/2}_m$ of GH, is associated with adding a unit of magnetic flux to 
an existing antisymmetric wave function
\begin{eqnarray}
\label{eq:vector2}
[z_1 z_2...z_k]^{k/2} &\equiv&  \left( \begin{array}{c} {[z_1 z_2 ...z_k]}^{k/2}_{k/2}~~~ \\  ~ \\{[z_1z_2 ... z_k]}^{k/2}_{k/2-1} \\ \vdots~~ \\ {[z_1z_2 ... z_k]}^{k/2}_{-k/2}~ \end{array} \right) \nonumber \\ ~ \nonumber \\
 {[z_1 z_2...z_k ]}^{k/2}_{k/2} &\equiv& {z_1 z_2 .... z_k \over  \sqrt{k!}} \nonumber \\
 {[z_1 z_2...z_k]}^{k/2}_{m-1} &\equiv& \left( {\partial_1} + {\partial_2} + ... + {\partial_k} \right) {[z_1 z_2 ... z_k]}^{k/2}_{m} \nonumber \\
 && 
\end{eqnarray}
where $\partial_i \equiv \partial/ \partial z_i$: the $N$-electron translation operator is now used as a lowering operator.
The vector components are the elementary symmetric polynomials for $N$ particles \cite{Schur}, e.g., 
 for $N$=4
\begin{eqnarray}
&&[z_1z_2z_3z_4]^2 = \nonumber \\
&&{1 \over \sqrt{4!}} \left( \begin{array}{c} z_1 z_2 z_3 z_4 \\ z_1 z_2 z_3 + z_1 z_2 z_4+ z_1 z_3 z_4 + z_2 z_3 z_4 \\  2! (z_1 z_2 +z_1 z_3+z_2 z_3+z_1 z_4+z_2 z_4 +z_3 z_4) \\ 3! (z_1+ z_2+z_3+z_4) \\ 4! \end{array} \right) \nonumber
\end{eqnarray}
A translationally invariant (scalar) quantity we will later use
\begin{equation}
 R_N(i) = \prod_{\begin{footnotesize} \begin{array}{c} j=1 \\ j \ne i \end{array} \end{footnotesize}} ^N(z_i-z_j)
 \label{eq:R}
\end{equation}
is a dot product of the two kinds of vectors,
\begin{eqnarray}
\label{eq:dot2}
 \left[ z_{1} \right]^{N-1 \over2} \odot \left[ z_2 z_3 ...z_N \right]^{N-1 \over 2}&=&(z_1-z_2)(z_1-z_3) ... (z_1-z_N)  \nonumber \\
 &=& R_N(1).
\end{eqnarray}

Finally, it is also possible to combine two planar spinors to form other spinors of a given rank,
analogous to spherical tensor products on the sphere.
From $[A]^{k_1}$ and $[B]^{k_2}$, spinors of rank $k_1$ and $k_2$ with $2k_1+1$ and
$2k_2+1$ components, respectively,  a new spinor of rank $k$ can be formed
that transforms properly under the planar  lowering (translation)  operator $\sum \partial_i$ 
\begin{eqnarray}
\left[[A]^{k_1} \otimes [B]^{k_2} \right]^{k}_m \equiv~~~~~~~~~~~~~~~~~~~~~~~~~  \nonumber \\
\left[ {(k-m)! \over (k+m)!} \right]^{1/2} \sum_{m_1=-k_1}^{k_1} 
\sum_{m_2=-k_2}^{k_2} \left[ {(k_1+m_1)! (k_2+m_2)! \over (k_1-m_1)! (k_2-m_2)! }\right]^{1/2}  \nonumber \\
\times \langle {k_1 } , m_1; {k_2}, m_2 | {k} ,m \rangle ~
[A]^{k_1}_{m_1} [B]^{k_2}_{m_2},~~~~~~\nonumber \\ ~~\nonumber \\
 -k \le m \le k~~~~~~~~~~~|k_1-k_2|\le k \le k_1+k_2~~~~~~~~~~
 \label{eq:tensor}
\end{eqnarray}
where the bracket is a Clebsch-Gordan coefficient.
In particular, our previously defined scalar product is
\begin{equation}
[A]^{k_1} \odot [B]^{k_1}  = { \sqrt{2 k_1+1} }~ \left[ [A]^{k_1} \otimes [B]^{k_1} ]\right]_0^0
\end{equation}
~\\

\noindent
{\bf Laughlin's Wave Function:} Laughlin constructed N-particle states $|N,m \rangle$ as approximate
variational ground states of fractional filling $1/m$, with $m$ odd to ensure antisymmetry.  On
the plane, designating the
coordinate-space form as $L[N,m,\{z_i,i=1,N\}] \equiv \langle \{z_i\} | N,m \rangle$, 
\begin{equation}
\label{eq:Laughlin}
   L[N,m,\{z_i\}] \sim \prod_{j>k=1}^N (z_j-z_k)^m e^{- \sum_i |z_i|^2/2 },
\end{equation}
where $\sim$ indicates we have defined this wave function up to normalization.  
Incrementing $m$ yields the fully filled ($m$=1), 1/3rd-filled ($m$=3), and 1/5th-filled ($m$=5) N-electron states.  
Although written in terms of $N$ single-electron coordinates, this wave function effectively depends only 
on $N-1$ intrinsic coordinates, as the center-of-mass associated with
the factor $\sum |z_i|^2$ is fixed in its lowest harmonic oscillator state, as one can show by transforming to Jacobi coordinates.

Up to normalization, the IQHE ($m=1$) state can be rewritten in several ways
\begin{eqnarray}
&&L[N,m=1] \sim \left| \begin{array}{cccc} z_1^{N-1} & z_2^{N-1}  & \cdots & z_N^{N-1} \\ z_1^{N-2} & z_2^{N-2} &  \cdots & z_N^{N-2} \\
     \vdots & \vdots & \ddots & \vdots \\
     1 & 1 & \cdots & 1 \end{array} \right| ~e^{- \sum_{i=1}^N |z_i|^2/2 }\nonumber \\
    & &~~~\sim  \sum_{q_i=-{N \over 2}+1}^{{N-1 \over 2} } \epsilon_{q_1, \cdots, q_N} [z_1]^{{N-1 \over2}}_{q_1} \cdots [z_N]^{{N-1 \over 2}}_{q_N} ~e^{- \sum_{i=1}^N |z_i|^2/2 }\nonumber \\
    &&~~~\sim~~~ \Big| [z_1]^{{N-1 \over 2} } \cdots [z_N]^{{N-1 \over 2}} \Big|~e^{- \sum_{i=1}^N |z_i|^2/2 }
\end{eqnarray}
The antisymmetric tensor $\epsilon_{q_1, \cdots, q_N}$ produces a scalar contraction on $N$ spinors, and thus can be regarded as
a generalization of the dot product between two vectors we defined previously.  The last form states that the
columns of the determinant can be taken to be the single-electron vectors we have formed. (Recall row normalization
is not relevant in a determinant.)

Laughlin's wave function can be written as the $m$th power of a determinant or alternatively as
a single determinant
\begin{eqnarray}
 \label{eq:Laughlin2}
 &&    L[N,m]   \sim \left| \begin{array}{cccc} z_1^{N-1} & z_2^{N-1}  & \cdots & z_N^{N-1} \\ z_1^{N-2} & z_2^{N-2} &  \cdots & z_N^{N-2} \\
     \vdots & \vdots & \ddots & \vdots \\
     1 & 1 & \cdots & 1 \end{array} \right|^m~e^{- \sum_{i=1}^N |z_i|^2/2 }  \nonumber \\
     && ~~ \nonumber \\
 &&~~\sim \Big| [z_1]^{{N-1 \over 2} } R^{m-1 \over 2}_N(1) ~~\cdots~~ [z_N]^{{N -1\over 2}} R^{m-1 \over 2} _N(N)\Big|  \nonumber \\
 &&~~~~~~~~~~~~~~~~~~~~~~~~~~~~~e^{- \sum_{i=1}^N |z_i|^2/2 }
 \end{eqnarray}
Even though the Laughlin wave function involves interacting electrons in a partially filled shell,
the second form above states that the wave function can be expressed in a 
closed-sub-shell or noninteracting form, mathematically analogous to the $\nu=1$ case, if the electrons are replaced by 
quasi-electrons.  This can be taken as the definition of the Laughlin quasi-electron, which for $m=3$ is
 \begin{eqnarray}
&& [z_1]^{{N-1 \over 2} } _{q} R_N(1)  \nonumber \\
&&~~ =  [z_1]^{{N-1 \over 2} }_q ~~ [z_1]^{{N-1 \over 2} } \odot \nonumber [z_2 \cdots z_N]^{N-1 \over 2}  \\
&&~~ \sim \left[ [z_1]^{N-1} \otimes [z_2 \cdots z_N]^{N-1 \over 2} \right]^{N-1 \over 2}_q 
\label{eq:quasiel}
 \end{eqnarray}
  
 The anti-aligned coupling in Eq. (\ref{eq:quasiel})  is favored energetically, 
 producing a factor of $(z_1-z_i)$, for
 all $i$.  The flux creation operator $[z_2 \cdots z_N]$ is symmetric among
 particle exchange -- the components are the elementary Schur polynomials for $N-1$ coordinates.  
 Thus the correlations that
 Laughlin builds in to his wave function treat all particles equivalently, even though in a given 
 configuration some particles $i$ will be closer to particle 1, and some farther away.   His construction respects 
 the scale invariance of the Coulomb potential -- classically, given a solution at one density, others
 could be obtain by a simple rescaling of the magnetic length.  For a system of quantum mechanical
 fermions, this rescaling
 is restricted to odd $m$.  The GH and GH$^2$ constructions are guided by very similar considerations.\\

\begin{figure}
\begin{center}
\includegraphics[width=0.46\textwidth]{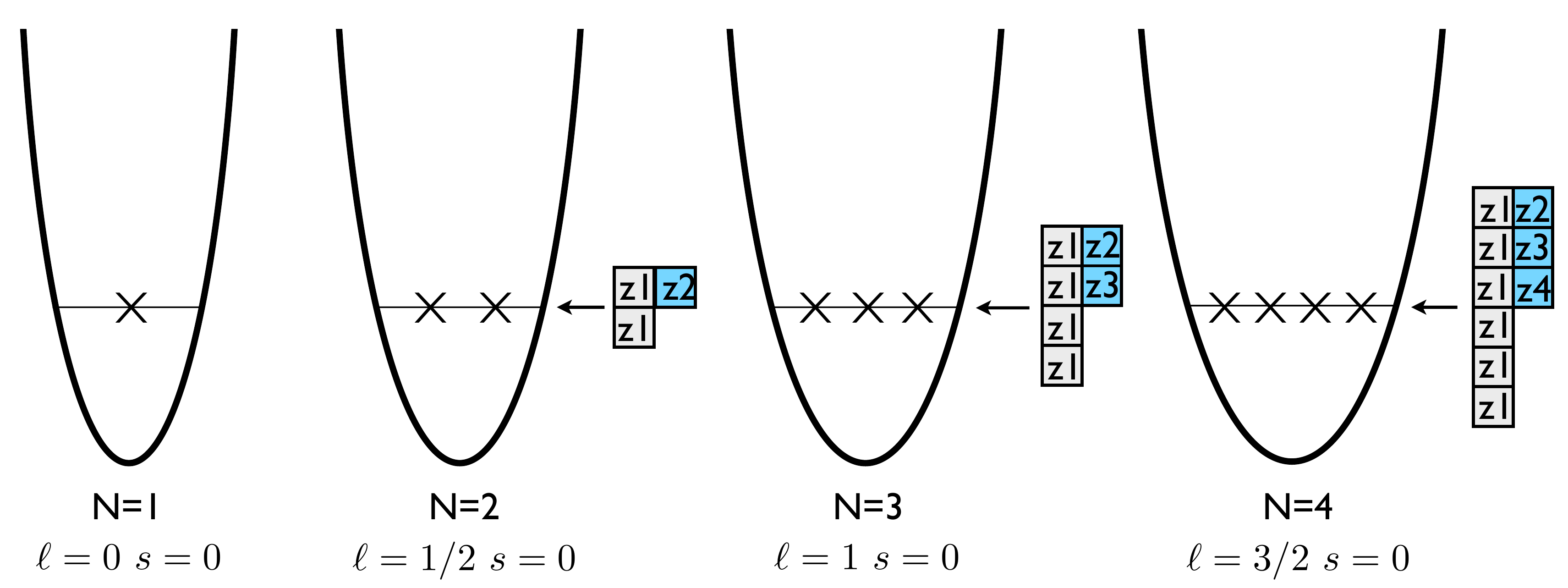}
\end{center}
\caption{Laughlin's 1/3rd-filled state as a function of electron number $N$, depicted as a quasi-electron
closed sub-shell.  The Laughlin quasi-electron is an anti-aligned product.}
\label{fig:Fig13}
\end{figure}

\noindent
{\bf Some properties of Laughlin wave function:} 
\noindent
Figure \ref{fig:Fig13} depicts Laughlin's $m=3$ wave function geometrically, as a closed sub-shell of quasi-electrons, 
as $N$ is incremented. \\

\noindent
{\it Recursion relation:}   For fixed $\nu$, with each increment in $N$, one electron and $m$ units of magnetic flux are
added overall to the wave function.
The scalar recursion operator describing this process acts on the $N$-particle wave to produce the
$(N+1)$-particle wave function,
\begin{eqnarray}
L[N+1,m]= R_{N+1}^m (N+1) e^{-|z_{N+1}|^2/2}  L[N,m] 
\label{eq:Lrecur}
\end{eqnarray}
where for m=3
\[ R_{N+1}^3(N+1)  =\left[ z_{N+1}\right]^{3 N \over 2} \odot \left[ z_1^3 z_2^3 ...z_N^3 \right]^{3 N \over 2} \]
and where
\[ [z_1^m \cdots z_N^m]^{N m \over 2}_{N m \over 2} \equiv {1 \over \sqrt{(mN)!}} z_1^m \cdots z_N^m \]
with the remaining components obtained by lowering, as described previously.
The operator is (necessarily) a scalar and symmetric among exchange of coordinates of the pre-existing electrons.  
In is easy to convert this recursion operator into its second quantized form, then employ it to recursively
generate the $N$-electron, starting from the trivial $N=1$ state. \\

\noindent
{\it Fractional charges:} Experiments have identified fractionally charged FQHE excitations \cite{expfc}, while theory 
can generate corresponding wave functions from recursion relations \cite{RMP,Halperin}.  Labeling the location of a 
hole by a coordinate $z_0$, one can define the wave functions 
\begin{eqnarray}
\Psi^\prime_q(z_0) = e^{-|z_0|^2/2} \times~~~~~~~~~~~~~~~~~~~~~~~~~~~~~~~~~~~~~ \nonumber \\
 \left\{ \begin{array}{cl}  \left[ z_0 \right]^{N \over 2} \odot \left[ z_1 z_2 \cdots z_N \right]^{N \over 2}  L[N,m] & q=-{e \over m} \\
~&~\\
\left[ z_0 \right]^{N}  \odot \left[ z_1^2 z_2^2 \cdots z_N^2 \right]^N L[N,m] & q=-{2e \over m} \\
\vdots & \\
\left[ z_0 \right]^{mN \over 2}  \odot \left[ z_1^m z_2^m \cdots z_N^m \right]^{m N \over 2} L[N,m] & q=-e \end{array} \right.
\label{eq:FC}
\end{eqnarray}
that insert a hole in all possible locations in the shell.
As the last step is identical to the recursion relation that creates the $N+1$-electron state, apart 
for the substitution $z_{N+1} \rightarrow z_0$, one
identifies the charge of the corresponding hole state as $q=-1$, relative to the $N+1$-electron state:
that charge would be cancelled by adding an electron.  The remaining states, as cycles that accumulate
to the last state, then carry the charges indicated.
Note that each component of the scalar products above corresponds to the insertion of a hole at a specific location,
breaking translational invariance, which is then restored when the dot-product sum is taken over all such components. 

A similar cycle of substitutions can be carried out, given a quasi-electron single-determinant form for the wave function.
For the case of $m=3$,  the determinant in Eq. (\ref{eq:Laughlin2}) evolves as following
\begin{small}
\begin{eqnarray}
 \Big| [z_1]^{{N-1 \over 2} } R_N(1) ~~\cdots~~ [z_N]^{{N -1\over 2}} R _N(N)\Big|  \rightarrow ~~~~~~~~~~\nonumber \\
  \Big| [z_1]^{{N+1 \over 2} } R_{N}(1) ~~\cdots~~ [z_N]^{{N +1\over 2}} R _{N}(N) ~~[z_0]^{N+1 \over 2} \Big| \rightarrow ~~~~~~\nonumber \\
    \Big| [z_1]^{{N+1 \over 2} } R^{z_0}_{N+1}(1) ~~\cdots~~ [z_N]^{{N +1\over 2}} R^{z_0} _{N+1}(N) ~~[z_0]^{N+1 \over 2} \Big| \rightarrow~~~ \nonumber \\
        \Big| [z_1]^{{N+1 \over 2} } R^{z_0}_{N+1}(1) ~~\cdots~~ [z_N]^{{N +1\over 2}} R ^{z_0}_{N+1}(N) ~~[z_0]^{N+1 \over 2}R _{N+1}(z_0) \Big|  \nonumber 
\end{eqnarray}
\end{small}
where the notation is
\begin{eqnarray}
 R_{N+1}^{z_0}(1)  &\equiv& \left[ z_{1}\right]^{ N \over 2} \odot \left[ z_2 z_3 ...z_N z_0 \right]^{ N \over 2} \nonumber \\
  R_{N+1}(z_0)  &\equiv& \left[ z_{0}\right]^{ N \over 2} \odot \left[ z_1 z_2 ...z_N \right]^{ N \over 2} \nonumber
\end{eqnarray}
Consequently, in the Laughlin case, it is sufficient to have either the recursion relation or the explicit single-determinant quasi-electron
form of the wave function to be able to introduce the defects that carry fractional charge.  We will present a general single-determinant
quasi-electron wave function for the FQHE later in this paper.\\

\begin{figure}
\begin{center}
\includegraphics[width=0.46\textwidth]{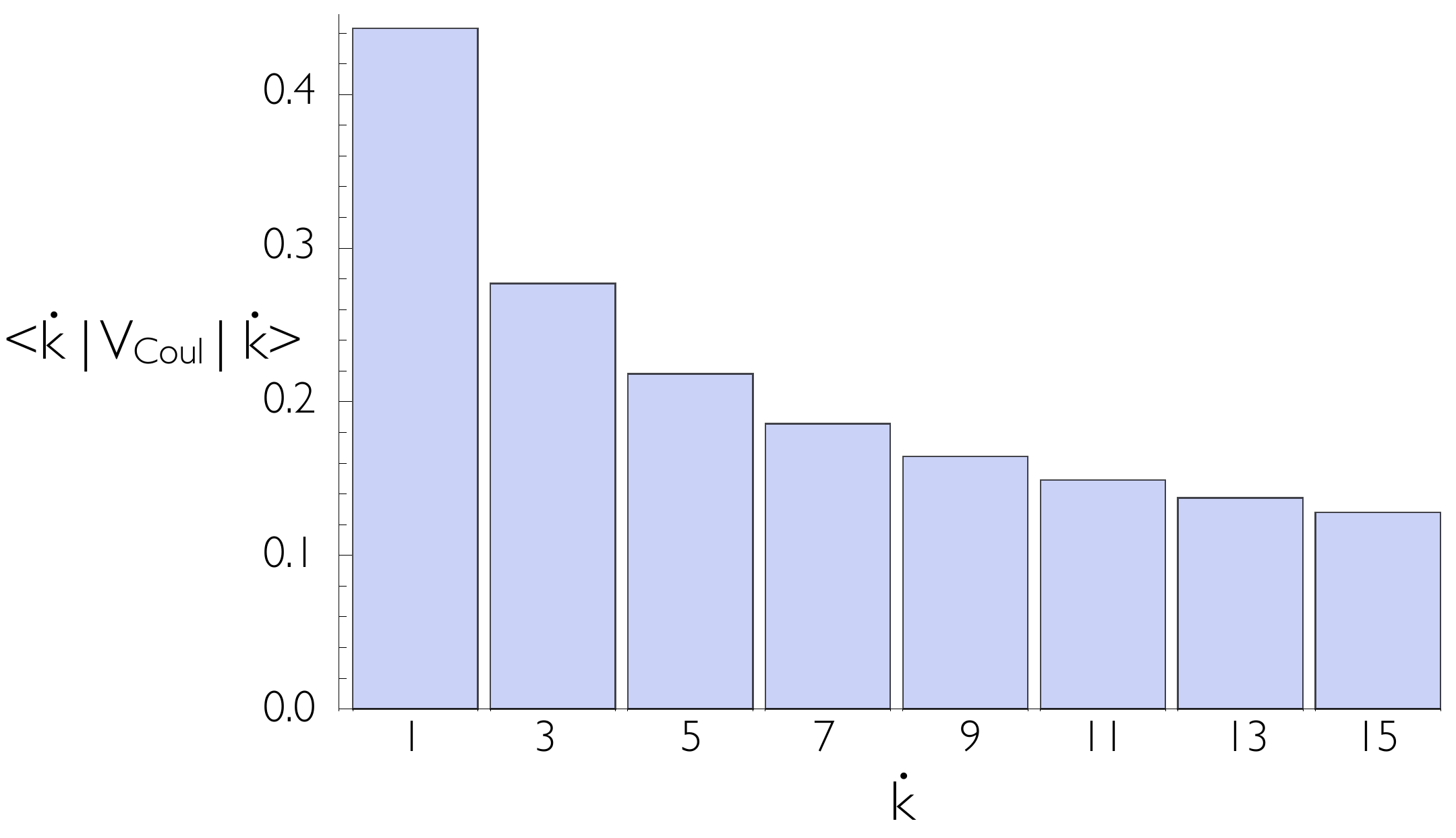}
\end{center}
\caption{Expectation value of Coulomb potential between the two electron relative wave function
$| \dot{k} \rangle$ in units of $\alpha (\hbar c/a_0)$.
\label{fig:Figure1}}
\end{figure}

\noindent 
{\it The pseudopotential:}  To evaluate the two-electron Coulomb matrix element it is convenient
to transform from state
vectors $| k_1 k_2 \rangle$ to the basis $| \dot{k}~ k_\mathrm{CM} \rangle$ involving
the Jacobi coordinates $\dot{z}=(z_1-z_2)/\sqrt{2}$ and $z_\mathrm{CM} = (z_1+z_2)/\sqrt{2}$.
The relative wave functions allowed by antisymmetry are
\begin{equation}
\label{eq:twop}
\langle \dot{z} | \dot{k} \rangle={1 \over \sqrt{ \pi \dot{k}!} }~\dot{z}^{\dot{k}} e^{-|\dot{z}|^2/2},~\dot{k}=1,3,5,...
\end{equation}
The nonzero matrix elements of the Coulomb potential,
\begin{equation}
\label{eq:Coulomb}
\langle \dot{k} | V_\mathrm{Coul} | \dot{k} \rangle =  \alpha \left( \hbar c \over a_0 \right) {1 \over 2^{\dot{k}+1} \dot{k}! } \sqrt{\pi} (2 \dot{k}-1)!!
\end{equation}
where $\alpha$ is the fine structure constant, are plotted in Fig.  \ref{fig:Figure1} as a function
of $\dot{k}$.  One consequence of Laughlin's construction is the exclusion of
the most repulsive  $\dot{k}$=1 (p-wave, $m$=3) and $\dot{k}=3$ (f-wave, $m$=5) components of the Coulomb
interaction.  The Laughlin wave functions can be viewed as a low-momentum restrictions of the true wave functions
that capture most of the physics relevant to long-wavelength properties of the system, while excluding high-frequency
components.

Haldane \cite{Haldane} intoduced an associated pseudo-potential for the FLL, e,g., 
\begin{eqnarray}
\label{eq:LEFT}
V_{1 \over 3} =V_0 \left[ {\overleftarrow{\partial} \over \partial \dot{z}^*} + {\dot{z} \over 2} \right]  \delta(\dot{z})
 \left[  {\overrightarrow{\partial} \over \partial \dot{z}} +{\dot{z}^* \over 2} \right]  
 = V_0~ {\overleftarrow{\partial} \over \partial \dot{z}^*} ~ \delta(\dot{z})~{\overrightarrow{\partial} \over \partial \dot{z}} \nonumber \\
 \langle \dot{k} | V_{1 \over 3} | \dot{k} \rangle = \delta_{\dot{k},1} {V_0 \over 2 \pi}~~~~~~~~~~~~~~~~~~~~~~~~~~~
\end{eqnarray}
This provides an order parameter for the Laughlin state, with
the density the control parameter: translationally invariant ground states with $\langle V_{1 \over 3} \rangle=0$ exist for $\nu \le 1/3$. 
This order is encoded in the form of the Laughlin quasi-electron, clearly.  Thus one anticipates that the quasi-electrons
relevant to other fractional fillings will have some connection to the Haldane pseudopotential, and that there
should be some change in the behavior of $\langle V_{1 \over 3} \rangle$ at the densities marking the relevant
fillings.

\section{III. Construction of the GH Operators}
\label{sec:GH}
\noindent
The GH operator construction was done in spherical geometry.
We retain that geometry here, making the transition to the plane at a later point.

Jain's general hierarchy wave function takes the form (borrowing the
$\ell s$ notation of GH)
\begin{eqnarray}
     \Phi^\mathrm{Jain}_{m,\ell,s}  &=& P_{FLL} \left( \Phi_{\ell,s}^\mathrm{Jain}  \left[ \prod_{i< j =1}^N u(i) ]\cdot u(j) \right]^{m-1} \right),  \nonumber \\
     &&~l \ge s,~~ s=0, 1/2, 1, ...~~~m=3,5, ...
 \label{eq:Jain}
\end{eqnarray}
where $\Phi_{\ell,s}^\mathrm{Jain}$ is the antisymmetric scalar operator obtained from the first $2s+1$ IQHE shells
filled by $(2 \ell+1)(2 s+1)$ electrons.   When the operator acts
on the symmetric bosonic state to the right,  
configurations involving highly excited magnetic states are generated.  Jain extracted a wave function for the FLL by numerically 
projecting out all higher LL components.  $P_{FLL}$ is this projection operator.

The GH wave functions have the form
\begin{equation}
     \Phi^\mathrm{GH}_{m,\ell,s}  = \Phi_{\ell,s}^\mathrm{GH}  \left[ \prod_{i< j =1}^N u(i) ]\cdot u(j) \right]^{m-1}  
     ~\begin{array}{l}  s=0, 1/2, 1, ... \\ m=3,5, ... \end{array}
 \label{eq:JainGH}
\end{equation}
where $\ell$ amd $s$ are unconstrained apart from $(2\ell+1)(2s+1)=N$.   The $l \ge s$ ($l \le s$)  operators
generate hierarchy (conjugate) states.\\

\noindent
{\it The GH Operator for $\nu$=2/5:}  We use the simplest non-Laughlin case of $\nu$=2/5 
to illustrate the GH operator construction.   The Laughlin state at $\nu=1/3$ is defined by its
two-electron correlation function.  As the density is increased above this value, local overdensities 
arise: quantum mechanics prevents the smooth rescaling of distances that would occur in
a degenerate classical system with a scaleless potential.  The process begins with the 
percolation of $p$-wave pairs: more precisely, the removal of quanta from a given two-electron
correlation creates an overdensity, including a nonzero amplitude for finding the electrons in
a relative $p$-wave.  One seeks a trial
variational wave function that keeps local overdensities separated to the extent possible,
thus preventing any larger, energetically more costly perturbations in local density.  The construction must be 
possible at all relevant $N$ and must
produce wave functions that are translational invariant and
homogeneous.  Laughlin addressed the same problem for point electrons as the ``overdensities", where the
available antisymmetric scalar correlations are odd powers of
\[ [u(1)] ^{1 \over 2}\odot [u(2)]^{1 \over 2} \equiv u(1)  \cdot u(2)  \]
$\Phi^\mathrm{Laughlin}_{m,\ell,s=0}$ is a product over all possible pairs of this form.

Identifying all candidate correlations among larger numbers of electrons is less trivial, though in the case of $\nu$=2/5
this can be done algebraically (see Appendix).  Alternatively, one might look
to Laughlin's $\nu=1/3$ wave function for guidance.   Laughlin's correction between pairs of electrons (1,2) and (3,4)
can be written as a scalar in multiple ways, e.g.,
\begin{eqnarray}
 &&u(1) \cdot u(2)~ u(1) \cdot u(3)~u(2) \cdot u(3)~ u(2) \cdot u(4) \nonumber \\
&&~~= -3 [u(1)^2 u(2)^2]^2 \odot [u(3)^2 u(4)^2]^2  \nonumber \\
&&~~~~~~~-2 \left( [u(1) u(2)]^1 \odot [u(3)u(4)]^1 \right)^2 
\label{eq:L}
\end{eqnarray}
The more elegant form of the Laughlin two-electron-to-two-electron correlation is
that given by the first line above, as the underlying structure of the wave function
is determined by the two-electron, not four-electron, correlation.  Yet the second expression
is helpful in identifying a scalar correlation important at higher densities.  Its 
spherical and planar forms are
\begin{eqnarray}
&&~[u(1) u(2)]^1 \odot [u(3)u(4)]^1 = u(1) \cdot u(3)~u(2) \cdot u(4)  \nonumber \\
&&~~~~~~~~~~~~~~~~~~~~~~~~~~~~~+ u(1) \cdot u(4)~ u(2)  \cdot u(3) \nonumber \\
&&~[z_1 z_2]^1 \odot [z_3 z_4]^1 = (z_1-z_3)(z_2-z_4)+(z_1-z_4)(z_2-z_3) \nonumber 
\label{eq:cortwo}
\end{eqnarray}
This scalar plays a role similar to  $u(1) \cdot u(2)$ in Laughlin's
wave function, operating among $N/2$ electron pairs.  Pairs so spaced clearly
correspond to a half-filled shell.  

The operator is symmetric under the interchanges $1 \leftrightarrow 2$
and $3 \leftrightarrow 4$, and thus would vanish under antisymmetrization. The 
antisymmetry can be restored by adding operators that destroy magnetic flux,
\[[u(1) u(2)]^1 \odot [u(3)u(4)]^1 ~ d(1) \cdot d(2)~ d(3) \cdot d(4).\]
As a scalar, $d(1) \cdot d(2)$ acting only on the relative wave function of any electron pair,
removing a quanta. 
Consequently it effectively acts at all inter-electron separations to bring the two electrons
closer together.  The ``defect" or over-density it creates includes at the shortest distance
scale a $p$-wave component.
Here $d_q=(-1)^{1/2+q} d/du_{-q}$ so that $d(1) \cdot d(2) u(1) \cdot u(2) =2$.
One can now antisymmetrize (it is sufficient to do so over the three distinct choices for
the clustering, $\{(12), (34) \}$, $\{(13),(24)\} $, and $\{ (14), (23)\}$,
\begin{eqnarray}
 \mathcal{A} \left[  [u(1) u(2)]^1 \odot [u(3)u(4)]^1~d(1) \cdot d(2)~ d(3) \cdot d(4)  \right] ~~~
\end{eqnarray} 
In the Appendix we show that this operator is associated with one of two symmetric
polynomials that provide a basis for all $N=4$ wave functions. 

The construction can be extended to any even $N$.  We introduce 
an index $I$ to denote a given pair of electrons, in some partition of the electrons, e.g., $\{I\} = \{ (1,2),(3,4), ...., (N-1,N) \}$, and define
\[ L^{s=1/2}_d(I=(1,2)) \equiv d(1) \cdot d(2) \]
\[ U[I=1] \odot U[I=2] \equiv [u(1)u(2)]^1 \odot [u(3)u(4)]^1 \]
\[\Phi^\mathrm{GH}_{\ell ,s=1/2} = \mathcal{A} \left[  \left( \displaystyle\prod_{I<J=2}^{N/2} U[I] \odot U[J] \right)\left( \displaystyle\prod_{I=1}^{N/2} L_d(I) \right) \right] \]
again with $N=(2 \ell+1)(2s+1)$, $2\ell=N/2+1$ when $s=1/2$.  At large $N$
the first term determines the filling, while the second produces the $N/2$ defects
that are then arranged in a manner that follows Laughlin's construction.
Antisymmetrization yields a simple determinant
\begin{eqnarray}
 \Phi^\mathrm{GH}_{\ell ,s=1/2} &=&
 \sum_{m_i=-\ell}^{\ell} ~\sum_{q_i=-1/2}^{1/2}  \epsilon_{M_1, \cdots, M_N} [u(1)]^{\ell} _{m_1} \cdots [u(N)]^ {\ell}_{m_N} \nonumber \\
&& \times [d(1)]^{1/2}_{q_1} \cdots [d(N)]^{1/2}_{q_N}
\label{eq:simpled}
\end{eqnarray}
where $M_i = (m_i,q_i)$ and
\begin{equation}
 [u]^{ \ell}_m \equiv \mathcal{D}^\ell_{\ell,m} = \left[ (2 \ell)! \over (\ell+m)!(\ell-m)! \right]^{1/2} u_{1/2}^{\ell+m} u_{-1/2}^{\ell-m} 
 \tag{\ref{eq:us}}
 \end{equation}
The $d$s can be written to the left or to the right of the $u$s, without changing the determinant.  This ``$\ell s=1/2$'' form of the operator --
the representation showing the correlation structure --
can be recast, in the usual way, in an equivalent $(\ell)j$ form, which is the quasi-electron representation that reveals the FLL sub-shell structure
induced by the correlations.
For $\nu=2/5$ and thus $s=1/2$,  $j= \ell \pm 1/2$, yielding
\begin{widetext}
  \begin{small}
  \begin{eqnarray} 
  && \Phi^\mathrm{GH}_{\ell , s=1/2}
   = \left| \begin{array}{lllll} ~\left[[u(1)]^{\ell} \otimes [d(1)]^{1 \over 2} \right]^{\ell+{1 \over 2}}& \left[ [u(2)]^{\ell} \otimes [d(2)]^{1 \over 2} \right]^{\ell+{1 \over 2}} & \cdots & \left[ [u(N-1)]^{\ell} \otimes [d(N-1)]^{1 \over 2}\right]^{\ell+{1 \over 2}} & \left[ [u(N)]^{\ell} \otimes [d(N)]^{1 \over 2} \right]^{\ell+{1 \over 2}}\\
 ~\left[  [u(1)]^{\ell} \otimes [d(1)]^{1 \over 2} \right]^{\ell-{1 \over 2}} & \left[ [u(2)]^{ \ell} \otimes [d(2)]^{1 \over 2} \right]^{\ell-{1 \over 2}}& \cdots & \left[ [u(N-1)]^{\ell} \otimes [d(N-1)]^{1 \over 2} \right]^{\ell-{1 \over 2}}& \left[ [u(N)]^{\ell} \otimes [d(N)]^{1 \over 2} \right]^{\ell-{1 \over 2}} \end{array}  \right| \nonumber \\
 &&= \mathcal{A} \left[ \sum_{m's=-(\ell+1/2)}^{\ell+1/2}   \epsilon_{m_1, \cdots, m_{2 \ell+2}} \left[\left [u\left({N/2}\right)\right]^\ell \otimes \left[d\left({N/2} \right)\right]^{1 \over 2} \right]^{l+1/2}_{m_1} \cdots \left[ \left[u \left(N\right)\right]^\ell \otimes \left[ d \left(N \right)\right]^{1 \over 2} \right]^{\ell+1/2}_{m_{2\ell+2} }\right. \nonumber \\
&&~~~~~~~~~~~\times \left. \sum_{m's=-(\ell-1/2)}^{\ell-1/2}   \epsilon_{m_1, \cdots, m_{2 \ell}} \left[ \left[u(1)\right]^\ell \otimes \left[ d(1) \right]^{1 \over 2}\right]^{\ell-1/2}_{m_1}\cdots \left[ \left[u \left({N/2} -1\right)\right]^\ell \otimes \left[ d \left({N/2}-1 \right)\right]^{1 \over 2}  \right]^{\ell-1/2}_{m_{2 \ell}} \right]
  \label{eq:52}
  \end{eqnarray}
  \end{small}
  \end{widetext}
  Each column of the antisymmetric $N \times N$ determinant is a direct sum 
  of the aligned and anti-aligned vectors.  The second, more explicit ``filled sub-shell" form consists of
  a $(N/2-1)$-electron lower ($j=\ell-1/2$) sub-shell and a $(N/2+1)$-electron upper ($j=\ell+1/2$) 
  sub-shell.    $\mathcal{A}$ antisymmetrizes over exchange of particles among the two closed
  shells.  This operator generates the $\nu=2/5$ state of the $m=3$ hierarchy, when it acts on the symmetric $m=2$ $N$-electron state,
  and the $\nu= 2/9$ state, when it acts on the $m=4$ symmetric state.  \\

  \noindent
  {\it $\nu$=2/3 case:} The $\nu=2/3$ conjugate state is the densest of the series ....4/7,3/5,2/3 and marks the filing
  where only isolated, two-electron droplets of the $\nu=1/3$ remain: we can create these under-dense regions
  by applying $N/2$ operators of the type $u(1) \cdot u(2)$ to the half-filled shell.  The operators that remove 
  flux from the half-filled shell to produce a state with $\nu=2/3$ are of the form $[d(1) d(2)]^1 \cdot [d(3) d(4)]^1$
  operating between all of the defects.  The arguments are the reverse of those for $\nu=2/5$.  We obtain for $N$=4
  \begin{equation}
 \mathcal{A} \left[ u(1) \cdot u(2)~ u(3) \cdot u(4) ~ [d(1) d(2)]^1 \odot [d(3)d(4)]^1~ \right]
 \end{equation}
  This operator and that for $N=4$ $\nu=2/5$ are identical after 
  antisymmetrization (a pattern that continues for all similar elementary hierarchy/conjugate pairs, and is associated
  with a symmetry at $\nu=1/2$ we discuss later).
  But for $N>4$ $s$ is incremented, while $\ell=1/2$ is fixed, yielding
  \begin{widetext}
  \begin{small}
  \begin{eqnarray} 
  && \Phi^\mathrm{GH}_{\ell=1/2 , s} 
  = \left| \begin{array}{lllll} ~\left[ [u(1)]^{1 \over 2} \otimes [d(1)]^{s} \right]^{s+{1 \over 2}}& \left[ [u(2)]^{1 \over 2} \otimes [d(2)]^{s} \right]^{s+{1 \over 2}} & \cdots &\left [u(N-1)]^{1 \over 2} \otimes [d(N-1)]^{s} \right]^{s+{1 \over 2}} & \left[ [u(N)]^{1 \over 2} \otimes [d(N)]^s \right]^{s+{1 \over 2}} \\
 ~ \left[ [u(1)]^{1 \over 2} \otimes [d(1)]^{s} \right]^{s-{1 \over 2}}&\left[  [u(2)]^{1 \over 2} \otimes [d(2)]^{s} \right]^{s-{1 \over 2}} & \cdots & \left[ [u(N-1)]^{1 \over 2} \otimes [d(N-1)]^{s} \right]^{s-{1 \over 2}} & \left[ [u(N)]^{1 \over 2} \otimes [d(N)]^{s} \right]^{s-{1 \over 2}} \\
\end{array}  \right| \nonumber \\
\label{eq:52a}
\end{eqnarray}
\end{small}
  \end{widetext}

  This can also be expressed as the product of two closed $(l=1/2 s)j=s\pm{1 \over 2}$ sub-shells, as in Eq. (\ref{eq:52}).
  The operator carries an effective filling $\nu=-1/2$.
   It generates the $\nu=2/3$ state of the $m=3$ hierarchy, when it acts on the symmetric $m=2$ $N$-electron
  state,
  and the $\nu= 2/7$ state, when it acts on the $m=4$ symmetric state.  \\

 \noindent
\textbf{Generalization for arbitrary $\ell,s$:}  The full set of GH operators is obtained by allowing $l$ and $s$
to vary over all integer and half-integer values, constrained by $(2l+1)(2s+1)=N$. The needed ``spreading operators"
\begin{eqnarray} 
[u(1)u(2)]^1 \cdot [u(3) u(4)]^1 \Rightarrow
U[I] \odot U[J] ~~~~~~~~~~~~~~ \nonumber \\
  \equiv  [u(1) \cdots u(2s+1)]^{s + {1 \over 2}} \odot [u(2s+2) \cdots u(4s+2)]^{s+{1 \over 2}} \nonumber 
\end{eqnarray}
operate between each pair $(I,J)$ of clusters.  This couples each electron $i$ in the first group to
a single electron $j$ in the second via a factor $u(i) \cdot u(j)$, symmetrized over all 
combinations, producing $1/(2s+1)$ of the pairs that would exist between the clusters in Laughlin's closed
shell. This is the effective filling.  For each cluster there is an operator that
destroys magnetic flux within that cluster, forming the defect
\begin{equation}
d(1) \cdot d(2) \Rightarrow 
 L_d^s(I) \equiv \prod_{i<j=2}^{2s+1} d(i) \cdot d(j), \nonumber
 \end{equation}
 while generating the needed antisymmetry.   When these operators are applied
 to the half-filled shell, the resulting wave functions include terms in which up
 to $N/(2s+1)$ droplets will appear, each containing $(2s+1)$ electrons that have some amplitude for being
 correlated as in the IQHE phase -- but no IQHE correlations among larger numbers of electrons.
 The evolution from  $\nu=1/3$ to  $\nu=1$ is marked by a series of steps in which progressively larger
 droplets of the IQHE phase appear, ending with a single $N$-electron droplet at $\nu=1$.
 
 Because the GH operators are scalars, the requirements of translational
 invariance and a homogeneous one-body density are satisfied.  The operators must
 be antisymmetrized among all the relevant partitions of electrons.  This
 yields the generalization of Eq. (\ref{eq:simpled})
\begin{eqnarray}
\label{eq:ls}
\Phi_{\ell, s}^\mathrm{GH} = ~~~~~~~~~~~~~~~~~~~~~~~~~~~~~~~~\nonumber \\
\sum_{m's,q's} \epsilon_{M_1 \cdots M_N} [u(1)]^{\ell}_{m_1} \cdots [u(N)]_{m_N}^{\ell}~ [d(1)]_{q_1}^{s} \cdots [d(N)]_{q_N}^{s},
\nonumber \\
~
\label{eq:GH3}
\end{eqnarray}
where $\epsilon$ is the antisymmetric tensor with $N$ indices and $M_i=(m_i,q_i)$, with $-\ell \le m_i \le \ell$ 
and $-s \le q_i \le s$.  The filling corresponding to Eq. (\ref{eq:Jain}) is then
\begin{equation}
 {N \over 2S+1} = { N \over (m-1)N+2\ell-2s -m+2} \nonumber \\
 \vspace*{0.2cm} ~\nonumber
 \end{equation}
Each distinct hierarchy state is indexed by some fixed $s$, with $\ell$ running over values $\ell>s$.  
$\ell \rightarrow \infty$ is the large $N$ limit, yielding
\begin{eqnarray}
 {N \over 2S+1}  \rightarrow {1 \over m-1 + {1 \over 2s+1}}, ~ s=0,1/2, 1,...  \nonumber 
\end{eqnarray}
producing the $m=3$ states with $\nu$=1/3, 2/5, 3/7, ...  Each conjugate state is indexed by some
fixed $\ell$, with $s \ge \ell$.  $s \rightarrow \infty$ is the large $N$ limit, yielding
\begin{eqnarray}
 {N \over 2S+1}  \rightarrow {1 \over m-1 - {1 \over 2 \ell+1}}, ~ \ell=0,1/2, 1,...  \nonumber 
\end{eqnarray}
producing the $m=3$ states with $\nu$ =1, 2/3, 3/5, 4/7,....  

The starting point for the GH$^2$ wave functions uses the GH operators in their $(\ell s)j$ forms.  Generalizing the previous result for $\nu=2/5$, 
\begin{widetext}
  \begin{small}
  \begin{eqnarray} 
  && \Phi^\mathrm{GH}_{\ell \ge  s}(\nu={1\over 2s+1})= \nonumber \\
  && \left| \begin{array}{lllll} ~\left[ \left[u(1)\right]^\ell \otimes \left[ d(1) \right]^s \right]^{\ell+s} &\left[ \left[ u(2) \right]^\ell \otimes \left[ d(2) \right]^s \right]^{\ell+s} & \cdots & \left[ \left[ u(N-1) \right]^\ell \otimes \left[ d(N-1) \right]^s \right]^{\ell+s} & \left[ \left[ u(N) \right]^\ell \otimes \left[ d (N) \right]^s \right]^{\ell+s}\\
  ~\left[ \left[u(1)\right]^\ell \otimes \left[ d(1) \right]^s \right]^{\ell+s-1} &\left[ \left[ u(2) \right]^\ell \otimes \left[ d(2) \right]^s \right]^{\ell+s-1} & \cdots & \left[ \left[ u(N-1) \right]^\ell \otimes \left[ d(N-1) \right]^s \right]^{\ell+s-1} & \left[ \left[ u(N) \right]^\ell \otimes \left[ d (N) \right]^s \right]^{\ell+s-1}\\
   ~~~~~~~~~~~\vdots & ~~~~~~~~~~\vdots &~ \vdots &~~~~~~~~~~~~~~~~~ \vdots &~~~~~~~~~~~ \vdots \\
          ~\left[ \left[u(1)\right]^\ell \otimes \left[ d(1) \right]^s \right]^{\ell-s} &\left[ \left[ u(2) \right]^\ell \otimes \left[ d(2) \right]^s \right]^{\ell-s} & \cdots & \left[ \left[ u(N-1) \right]^\ell \otimes \left[ d(N-1) \right]^s \right]^{\ell-s} & \left[ \left[ u(N) \right]^\ell \otimes \left[ d (N) \right]^s \right]^{\ell-s}  \end{array}  \right| 
  \label{eq:GHgen}
  \end{eqnarray}
  \end{small}
  \end{widetext}
 One can also rewrite this as a product of $2s+1$ closed-sub-shell operators, just as was done previously in Eqs. (\ref{eq:52}) and (\ref{eq:52a}).
 The GH$^2$ procedures imprint this sub-shell structure on the quasi-electrons, as described in the next section.\\
  
  \section{IV. Quasi-electrons and the GH$^2$ wave function}
  In this section we describe an alternative use of the GH operators that leads to a simple quasi-electron 
  description of the wave functions.
 Equation (\ref{eq:quasiel}) can be rewritten for the sphere as
  \begin{eqnarray}
 && R_N(i)= \prod_{ j\ne i=1}^N  u(i) \cdot u(j)  \nonumber \\
 && \mathrm{or}~R_N(N) \sim [u(N)]^{N-1 \over 2} \odot [u(1) \cdots u(N-1)]^{N-1 \over 2} \nonumber
 \end{eqnarray}
  An even power of a closed shell can be written
  \begin{eqnarray}
   L[N,m=2]&=& R_N(N) R_N(N-1) \cdots R_N(1) \nonumber \\
   L[N,m=4]&=& R_N^{2}(N) R_N^{2}(N-1) \cdots R_N^{2}(1) \nonumber
   \end{eqnarray}
Consequently the quasi-electron form of Laughlin's state on the sphere is, following Eq. (\ref{eq:Laughlin2}),
the determinant
    \begin{eqnarray}
   L[N,m=3]   \equiv
   \Big|  \left[  u (1) \right]^{N-1 \over 2} R_N(1)  \cdots  \left[ u (N) \right]^{N-1 \over 2} R_N(N) \Big|  \nonumber \\
  =  \Big| 
   \left[ \left[u(1)\right]^{N-1} \otimes \left[u(2) \cdots u(N)\right]^{N-1 \over 2}  \right]^{{N-1} \over 2}~~~~~~~~~~~~~~\nonumber \\
   ~~~~\cdots~~ \left[ \left[ u(N) \right]^{{N-1}} \otimes \left[u(1) \cdots u(N-1) \right]^{N-1 \over 2}  \right]^{N-1 \over 2}  \Big|~~~~~~
   \label{eq:form}
 \end{eqnarray}
 
 The resulting wave function is exactly equivalent to that obtained by acting with the Laughlin closed-shell
 operator on the half-filled shell.  But these two alternate procedures are not equivalent for $s \ne 0$
 GH operators.  Consequently there are two GH generalizations of Laughlin's construction, both of which 
 generate translational invariant, homogeneous states.  They differ because the derivatives creating the 
 defects act in slightly different ways in the two procedures.
 By using the GH operators as in Eq. (\ref{eq:form}), simpler wave functions revealing more of the underlying physics are obtained,
 \begin{widetext}
\begin{eqnarray}
&&\Phi^\mathrm{GH^2}_{m=3;\ell, s} = \nonumber \\
&&\left| \begin{array}{cccc}
   ~\left[ \left[u(1)\right]^\ell \otimes \left[ d(1)\right]^s \right]^{\ell+{s}}~R_N(1) ~~&~~\left[ \left[u(2)\right]^\ell \otimes \left[ d(2)\right]^s \right]^{\ell+{s}}~R_N(2) ~~&~~ \cdots ~~&~~ \left[ \left[u(N)\right]^\ell  \otimes \left[ d(N) \right]^s \right]^{\ell + {s}}~R_N(N) \\
      ~& ~ & ~& ~ \\
      \vdots & \vdots & \vdots & \vdots \\
~&~&~&~\\
   ~\left[ \left[u(1)\right]^\ell \otimes \left[ d(1)\right]^s \right]^{\ell-{s}}~R_N(1) ~~&~~\left[ \left[u(2)\right]^\ell \otimes \left[ d(2)\right]^s \right]^{\ell-{s}}~R_N(2) ~~&~~ \cdots ~~&~~ \left[ \left[u(N)\right]^\ell  \otimes \left[ d(N) \right]^s \right]^{\ell - {s}}~R_N(N)
 \end{array} \right| ~~~~
\end{eqnarray}
Each column consists of $2s+1$ vectors and has a total length of 
$ \sum_{j=\ell-s}^{\ell+s} (2j+ 1) =(2\ell+1)(2 s+1) = N$. 
The derivatives can be evaluated from results in \cite{GH}, to yield the generalized quasi-electrons
\begin{eqnarray}
\big[ \left[ u(1) \right]^{\ell} \otimes \left[ d(1) \right]^{s} \big]^j ~R_N(1) &\sim&  \big[u^{N-1+2(\ell-s)}(1) \otimes [u(2) \cdots u(N)]^{N-1 \over 2} \big]^j  \nonumber \\
&\equiv&  \big[ [u(1)]^{{N-1 \over 2}+( \ell-s)} \otimes [u(2) \cdots u(N)]^{N-1 \over 2} \big]^j  \mathrm{~~with~~}
 \ell-s \le j \le \ell+s 
 \label{eq:derivative}
\end{eqnarray}
where the $\sim$ indicates we have ignored irrelevant factors that normalize the expression.  Thus we can write a 
general first hierarchy wave function as a simple set of closed sub-shells.
\begin{eqnarray}
&&\Phi^\mathrm{GH^2}_{m=3;\ell  \ge s}(\nu={2s+1 \over 4s+3}) =\nonumber \\
&&~~~\left| \begin{array}{lcl}
   \left[ \left[u(1)\right]^{{N-1 \over 2} +(\ell-s)} \otimes \left[u(2) \cdots u(N)\right]^{N-1 \over 2}  \right]^{\ell+s}~&~ \cdots ~&~\left[ \left[ u(N) \right]^{{N-1 \over 2}+(\ell-s)} \otimes \left[u(1) \cdots u(N-1) \right]^{N-1 \over 2}  \right]^{\ell+s}  \\
  ~~~~~~~~~~~~~~~~~~~~ \vdots ~& \vdots &~~~~~~~~~~~~~~~~~~~~ \vdots  \\
    \left[ \left[u(1)\right]^{{N-1 \over 2}+(\ell-s)} \otimes \left[ u(2) \cdots u(N) \right]^{N-1 \over 2}  \right]^{\ell-s}~&~ \cdots ~&~\left[ \left[u(N)\right]^{{N-1 \over 2}+(\ell-s)}\otimes \left[u(1) \cdots u(N-1) \right]^{N-1 \over 2}  \right]^{\ell-s} \end{array} \right| 
\end{eqnarray}
With the elimination of derivatives, we can now use our previous mapping of spherical tensor products
to planar tensor products, to find translationally invariant and homogeneous wave functions for the plane,
\begin{eqnarray}
&&\Phi^\mathrm{GH^2}_{m=3;\ell \ge  s}(\nu={2s+1 \over 4s+3}) = \left| \begin{array}{lcl}
   \left[ \left[z_1\right]^{{N-1 \over 2} +(\ell-s)} \otimes \left[z_2 \cdots z_N\right]^{N-1 \over 2}  \right]^{\ell+s}~&~ \cdots ~&~\left[ \left[ z_N \right]^{{N-1 \over 2}+(\ell-s)} \otimes \left[z_1 \cdots z_{N-1} \right]^{N-1 \over 2}  \right]^{\ell+s}  \\
  ~~~~~~~~~~~~~~~~~~~~ \vdots ~& \vdots &~~~~~~~~~~~~~~~~~~~~ \vdots  \\
    \left[ \left[z_1 \right]^{{N-1 \over 2}+(\ell-s)} \otimes \left[z_2 \cdots z_N \right]^{N-1 \over 2}  \right]^{\ell-s}~&~ \cdots ~&~\left[ \left[z_N \right]^{{N-1 \over 2}+(\ell-s)}\otimes \left[z_1 \cdots z_{N-1}  \right]^{N-1 \over 2}  \right]^{\ell-s} \end{array} \right|\nonumber \\
   && ~~~~~~~~~~~~~~~~~~~~~~~~~~~~~~~~~~~~~~~~~~~~~~~~~~~~~~~~~~~ \times e^{-\sum_{i=1}^N |z_i|^2/2}
\end{eqnarray}
where the planar vectors and their tensor product are defined in Eqs. (\ref{eq:vector}), (\ref{eq:vector2}), and
(\ref{eq:tensor}).   As was done in Eq. (\ref{eq:52}), this result can be rewritten as the product of $2s+1$ closed-shell wave functions, antisymmetrized
over all partitions of the electrons among the shells.

Similarly the conjugate states, indexed by $\ell$ with $s \ge \ell$, become
\begin{eqnarray}
&&\Phi^\mathrm{GH^2}_{m=3;\ell \le s}(\nu={2\ell+1 \over 4 \ell+1}) =\nonumber \\
&&~~~\left| \begin{array}{lcl}
   \left[ \left[u(1)\right]^{{N-1 \over 2} -(s-\ell)} \otimes \left[u(2) \cdots u(N)\right]^{N-1 \over 2}  \right]^{s+\ell}~&~ \cdots ~&~\left[ \left[ u(N) \right]^{{N-1 \over 2}-(s-\ell)} \otimes \left[u(1) \cdots u(N-1) \right]^{N-1 \over 2}  \right]^{s+\ell}  \\
  ~~~~~~~~~~~~~~~~~~~~ \vdots ~& \vdots &~~~~~~~~~~~~~~~~~~~~ \vdots  \\
    \left[ \left[u(1)\right]^{{N-1 \over 2}-(s-\ell)} \otimes \left[ u(2) \cdots u(N) \right]^{N-1 \over 2}  \right]^{s-\ell}~&~ \cdots ~&~\left[ \left[u(N)\right]^{{N-1 \over 2}-(s-\ell)}\otimes \left[u(1) \cdots u(N-1) \right]^{N-1 \over 2}  \right]^{s-\ell} \end{array} \right| 
\end{eqnarray}
\begin{eqnarray}
&&\Phi^\mathrm{GH^2}_{m=3;\ell  \le s}(\nu={2 \ell+1 \over 4 \ell+1}) =\left| \begin{array}{lcl}
   \left[ \left[z_1\right]^{{N-1 \over 2} -(s-\ell)} \otimes \left[z_2 \cdots z_N\right]^{N-1 \over 2}  \right]^{s+\ell}~&~ \cdots ~&~\left[ \left[ z_N \right]^{{N-1 \over 2}-(s-\ell)} \otimes \left[z_1 \cdots z_{N-1} \right]^{N-1 \over 2}  \right]^{s+\ell}  \\
  ~~~~~~~~~~~~~~~~~~~~ \vdots ~& \vdots &~~~~~~~~~~~~~~~~~~~~ \vdots  \\
    \left[ \left[z_1 \right]^{{N-1 \over 2}-(s-\ell)} \otimes \left[z_2 \cdots z_N \right]^{N-1 \over 2}  \right]^{s-\ell}~&~ \cdots ~&~\left[ \left[z_N \right]^{{N-1 \over 2}-(s-\ell)}\otimes \left[z_1 \cdots z_{N-1}  \right]^{N-1 \over 2}  \right]^{s-\ell} \end{array} \right| \nonumber \\
    && ~~~~~~~~~~~~~~~~~~~~~~~~~~~~~~~~~~~~~~~~~~~~~~~~~~~~~~~~~~~ \times e^{-\sum_{i=1}^N |z_i|^2/2}
\end{eqnarray}
Note that for $\nu=1$ ($\ell=0$) we encounter
another expression for the $N$-electron IQHE state as an $N \times N$ determinant
\begin{eqnarray}
&&\Phi^\mathrm{GH^2}_{m=3;\ell=0~ s={N-1 \over2} }(\nu=1) =\left| \begin{array}{lcl}
    \left[z_2 \cdots z_N\right]^{N-1 \over 2} ~&~ \cdots ~&~ \left[z_1 \cdots z_{N-1} \right]^{N-1 \over 2}   \end{array} \right| = \left| \begin{array}{ccc} [z_1]^{N-1 \over 2}  & \cdots & [z_N]^{N-1 \over 2}  \end{array} \right|
    ~~
\end{eqnarray}

The evolution of the sub-shell structure (the number of shells, their occupancies) of the states is quite interesting, as the angular momentum of the quasi-electrons
changes as new electrons are added.  This is more easily described geometrically.  We do so below. 
\end{widetext}

\subsection{Quasi-electron structure}
The results above show that the quasi-particles that arise in mapping the FQHE
hierarchy states of filling $p/(2p+1)$ into a noninteracting form 
are simple objects that carry the quantum numbers $N,L,L_z,$ and $\mathcal{I}$,
where the allowed $N$s are divisible by $p$, $L=\ell-s+\mathcal{I}-1={1 \over 2}({N \over p} -p)+\mathcal{I}-1$, and
$1 \le \mathcal{I} \le p$.   In addition to its role in the construction of translationally invariant and homogeneous
many-electron states, angular momentum is associated with variational strategies to minimize the Coulomb interaction, and
thus with the quantum number $\mathcal{I}$.  The generators of rotations \cite{Haldane} on the sphere
\[ L_{1m} = {\hbar \over 2}  \sum_i \left[ u(i) \otimes d(i) \right]^1_M \]
involve sums over ``pair-breaking" operators:  $[u(i) \otimes d(i)]^1$, acting on the a scalar pair $u(i) \cdot u(j)$, produces the aligned pair
$[u(i) \otimes u(j)]^1_m$ \cite{GH}.  That is, the signature of the correlations that arise with increasing density
is the generation of angular momentum.  Constructions 
like GH$^2$, in which regions of overdensity are kept separated to the extent possible by building in multi-electron
correlations, will natural lead to a tower of angular momentum sub-shells, and to a noninteracting quasi-electron 
picture in which the quasi-electrons fill the sub-shells of lowest $\mathcal{I}.$

The angular momentum connections can be made clearer by recasting the $m=3$ hierarchy $(\ell,s)$-labeled quasi-electrons 
into a form that employs the more familiar variables $N$ and magnetic flux (in elementary units) $2S$: $S$ also determines the number of single-electron states in the Hilbert
space, $2S+1$.   As $S=N-1+\ell-s$, the general form of the hierarchy  quasi-electron is
\begin{eqnarray}
\left[ [z_1]^{S-{N-1 \over 2}} \otimes [z_2 \cdots z_N]^{N-1 \over 2} \right]^{S-N+\mathcal{I}} ~ e^{-|z_1|^2/2}  \nonumber \\
1 \le \mathcal{I} \le 2s+1 \equiv p~~~~~~\nu < 1/2~~~~~~~~~~~~~~~~
\label{eq:qe2}
\end{eqnarray}
where $\mathcal{I}=1,2, \dots$ is the quasi-electron sub-shell index.  Quasi-electrons can be defined for any
$S$ and $N$, while their closed-sub-shell configurations arise only for certain values of $S$ and $N$.   For $m=3$ hierarchy 
states, those values are
\begin{eqnarray}
 S= N \left({4s+3 \over 4s+2}\right) -(s+3/2)~~~~~~~~~~~~ \nonumber \\
  \Rightarrow  \nu={N \over 2S+1} \rightarrow { 2 s+1 \over 4 s+3} ={p \over 2p+1} = {1 \over 3}, {2 \over 5}, {3 \over 7}, \cdots \nonumber
 \end{eqnarray}
with $N$ divisible by $2s+1$. 

 For $\mathcal{I}=1$ (the anti-aligned Laughlin case) one has
\begin{eqnarray}
 \left[ [z_1]^{S - {N-1 \over 2}} \otimes [z_2 \cdots z_N]^{N-1 \over 2} \right]^{S-N+1} \sim \nonumber \\
~  [z_1]^{S-N+1} ~ [z_1]^{N-1 \over 2} \odot  [z_2 \cdots z_N]^{N-1 \over 2} 
\label{eq:dc0}
  \end{eqnarray}
The second term is the product of the $(z_1-z_i)$ for all $i \in \{2, \cdots, N \}$.  Thus there are no
broken scalar pairs in the first sub-shell. Consequent one is guaranteed that the minimum number of quanta in the two-electron
relative wave function for electrons 1 and 2 is $(z_1-z_2)^3$, with one factor coming from quasi-electron 1,
one from quasi-electron 2, and one from the determinant.   

The $\mathcal{I}=2$ quasi-electron can be rewritten
 \begin{eqnarray}
\left[ [z_1]^{S-{N-1 \over 2}} \otimes [z_2 \cdots z_N]^{N-1 \over 2} \right]^{S-N+2} \sim~~~~ \nonumber \\
~\sum_{i=2}^N [z_1^{2S-2N+3}z_i]^{S-N+2} ~ [z_1]^{N-2  \over 2} \odot  [z_2 \cdots z_N;\bar{z}_i]^{N-2 \over 2}   \nonumber \\
~~
\label{eq:dc1}
\end{eqnarray}
Here 
$[z_1^{2S-2N+3}z_i]^{S-N+2}$ is the aligned coupling of $z_1^{2S-2N+3}$ and $z_i$, while
$[z_2 \cdots z_N;\bar{z}_i]^{N-2 \over 2}$ denotes the $(N-2)$-electron
Schur polynomial for the indicate coordinates with $z_i$ removed from the set.  Relative to the $\mathcal{I}=1$ quasi-electron,
the $\mathcal{I}=2$ quasi-electron is missing one elementary scalar $[z_1]^{1/2} \odot [z_i]^{1/2}$, 
replaced by $[z_1z_i]^1$.  The result of the breaking on the scalar pair is the generation of a new
quasi-electron with one additional full unit of angular momentum.   Consequently there will be terms in the wave function
where there is only one quantum $z_1-z_2$ in the 1-2 relative wave function.  The GH$^2$ $\nu=2/5$
wave function, corresponding to the filling of sub-shells $\mathcal{I}=$ 1 and 2, contains terms with $N/2$ $p$-wave correlations,
but has no component with three electrons in a relative $p$ wave.  If one isolates the term with $N/2$ p-wave interactions,
one finds the defects arranged as in Laughlin's $1/5$ state.  This is consistent with the identification of the filling of the state as $\nu=2/5$.

Similarly the $\mathcal{I}=3$ quasi-electron is
 \begin{eqnarray}
\left[ [z_1]^{S-{N-1 \over 2}} \otimes [z_2 \cdots z_N]^{N-1 \over 2} \right]^{S-N+3} \sim~~~~~~~~~ \nonumber \\
~\sum_{i>j=2}^N [z_1^{2S-2N+4}z_i z_j]^{S-N+3} ~ [z_1]^{N-3  \over 2} \odot  [z_2 \cdots z_N;\bar{z}_i \bar{z}_j ]^{N-3 \over 2}   \nonumber \\
~~
\label{eq:dc2}
\end{eqnarray}
It allows the alignment of three electrons, $[z_1 z_i z_j]^{3/2}$.  Consequently local 3-electron over-densities correlated
as in the $\nu=1$ wave function exist, but again the cost in energy of such fluctuations are minimized by isolating
the over-densities.  The GH$^2$ $\nu=3/7$ states (filled $\mathcal{I}$=1,2,3 sub-shells) contain components with $N/3$
droplets of the integer phase -- but with the centers of these droplets distributed in the plane like the
electrons in Laughlin's $\nu=1/7$ state.  The pattern continues for the remaining hierarchy states, with
larger numbers of sub-shells and progressively more broken elementary scalars.

The quasi-electron coupling to its neighbors is always through
the symmetric Schur-polynomial vector $[z_2 \cdots z_N]^{N-1 \over 2}$.  This is the reflection
of the scale invariance we have described before: with each successive increment in $\mathcal{I}$,
an additional quanta is removed from the coupling of electron 1 to its neighbors, without regard for the specific positions
of the neighbors.  This is the best approximation allowed in quantum mechanics to a uniform rescaling of
inter-electron distances.

The conjugate ($\nu \ge 1/2$) analog of Eq. (\ref{eq:qe2}) can be written in the same form, but with a 
different restriction on $\mathcal{I}$,
\begin{eqnarray}
&&\left[ [z_1]^{S-{N-1 \over 2}} \otimes [z_2 \cdots z_N]^{N-1 \over 2} \right]^{S-N+\mathcal{I}} ~ e^{-|z_1|^2/2} \nonumber \\
&&~~~{N \over 2 \ell+1} - 2 \ell \le \mathcal{I} \le  {N \over 2 \ell+1}~~~~~~\nu > 1/2 \nonumber
\label{eq:qe3}
\end{eqnarray}
where $\mathcal{I}-1$ continues to correspond to the number
of missing $z_1-z_i$ scalar couplings.  The restriction can also be written
\[ 2(s-\ell) +1\le \mathcal{I} \le 2s+1 \]
 As $s \ge \ell$ for conjugate states, we see that it is possible to construct an $\mathcal{I}=1$ 
 (Laughlin) quasi-electron only if the state is self-conjugate ($s=\ell$).  In such cases $S=N-1$, so this 
sub-shell carries an angular momentum of $S-N+\mathcal{I}=0$ and thus contains only a single
quasi-electron.   We will identify this sub-shell with the ``base" of a deep quasi-electron Fermi sea that
forms at $\nu=1/2$, in subsequent discussions.

The nondegenerate states corresponding to configurations of quasi-electrons in filled sub-shells are obtained for values
of the magnetic field $S$ and particle number $N$ satisfying
\begin{eqnarray}
 S= N \left({4l+1 \over 4l+2}\right) +(l-1/2)~~~~~ \nonumber \\
 \Rightarrow  \nu={N \over 2S+1} \rightarrow { 2\ell+1 \over 4 \ell+1} = 1, {2 \over 3}, {3 \over 5}, \cdots \nonumber
 \end{eqnarray}
with $N$ divisible by $2l+1$.  For fixed $\nu$
(fixed $\ell$) the number of occupied sub-shells is $2\ell+1$, but for every increment of the electron number $N$
by $2\ell+1$, the indices of the sub-shells $\mathcal{I}$ increase by one, indicating the number of electrons $i$
lacking a factor $z_1-z_i$ has increased by one in each sub-shell.  A simple example is the $\ell=0$, $\nu=1$
case, where the quasi-electron for electron one is $[z_2 \cdots z_N]$: there are no scalars $z_1-z_i$,
and thus the number of missing scalars is $N-1$.  The index of the occupied sub-shell is $\mathcal{I}=N$,
tracking the electron number.

For conjugate states of any filling, the quasi-electron of the lowest occupied sub-shell 
is an anti-aligned product of the two vectors: the missing scalars come not (as in the hierarchy case) from a non-anti-aligned
vector product, but from the fact that the length of $[z_2 \cdots z_N]$ exceeds that of the
spinor for electron 1, so that some $z_1-z_i$ elementary scalars must be missing. One finds
\begin{eqnarray}
\left[ [z_1]^{S-{N-1 \over 2}} \otimes [z_2 \cdots z_N]^{N-1 \over 2} \right]^{S-N+\mathcal{I}_\mathrm{min}} \sim ~~~~   \nonumber \\
\sum_{2 \le i_1< \cdots <i_{2 (s-\ell)}} ^N ~[\bar{z}_{i_1} \cdots \bar{z}_{i_{2 (s-\ell)}}]^{s-\ell}~~~~~~~~~~~~~\nonumber \\
\times  [z_1]^{{N-1 \over 2}-(s-\ell)}   \odot  [z_2 \cdots z_N; \bar{z}_{i_1} \cdots \bar{z}_{i_{2 \kappa}}]^{{N-1 \over 2}-(s-\ell)} 
\label{eq:nubig}
\end{eqnarray}
again yielding $\mathcal{I}_\mathrm{min} = 2(s-\ell)+1$.

The connections described here between clustering of planar wave functions, their representations  in terms of SU(2) algebra, and the vanishing of the bosonic
part of the wave function (the wave function with one power of the antisymmetric closed shell removed) when $\mathcal{I}+1$ electrons 
are placed at one point,  have been noticed before, most prominently by 
Stone and collaborators \cite{Stone}, who also explored the implications of such properties to global topological order and to the
nonabelian statistics of the quasi-particles.  A specific construction of the sort provided here, however, is new, as is the recognition
that these properties were implicit in the GH operators.  We will later 
use the results above to put the wave functions in their generalized Pfaffian form,  the most symmetric form of the GH$^2$ wave
function in which the composite fermions can be regarded as fermionic excitations of a bosonic vacuum state.\\

\noindent
{\it Relation to the pseudo-potential:} The ground-state expectation of Haldane's pseudopotential provides an order parameter
for Laughlin's state, and thus that one might expect other hierarchy states to mark densities where
similar transitions in this parameter occur.
While one can calculate this expectation value from GH$^2$ wave functions (this will be done in the future), a simpler surrogate
is to count the missing scalar pairs $z_i-z_j$, per electron and relative to the Laughlin
state, at each value of $\nu$.   This yields, in the large $N$ limit, 0 ($\nu=1/3$), 1/2 ($\nu=2/4$), 1 ($\nu=3/7$), 3/2 ($\nu=4/9$), ....
This implies a change in the slope of the expectation value of the Haldane pseudo-potential at the fillings marking the closed-shell states.

One can calculate the additional energy associated with the addition or removal of one quasi-electron at  these filings.  This 
addition/removal does not create a simple particle or hole state, as changing the electron
number alters the angular momentum of the existing quasi-electrons.  This process is illustrated in Fig. \ref{fig:FixedMag}.\\

\noindent
{\it Effective B field:}  The usual argument \cite{JainCF} leading to the conclusion that a CF is an electron that has absorbed two 
external magnetic flux units is based
on producing the reduced effective B field needed to account for the angular momentum and closed-shell structure
of the CFs.   Two flux units per electron are indeed absorbed into internal (scalar) quasi-electron wave functions of GH$^2$, although the 
GH$^2$ quasi-electron differs from Jain's picture, as it involves the coupling of an electron spinor to a single magnetic
flux spinor.   It may be helpful follow the flux associated with the addition of an electron to the wave function, to explain
the ``bookkeeping."

Quasi-electron labels $\ell$ and $s$ are related to the electron number $N$ and total magnetic flux through the sphere $2S$ by $S=N+\ell -s-1$.  Consider
quasi-electron 1 in the $\mathcal{I}=1$ (Laughlin) shell of a FQHE hierarchy state of filling $\nu$.   In
the mapping to quasi-electrons, $N-1$ flux units from electron 1 are used by (absorbed into) electrons 2 through $N$, screening
them from electron 1:
this is represented mathematically as an anti-aligned coupling of an angular-momentum-1/2 flux unit from electron 1
with the spinors of electrons 2 through $N$, 
forming a scalar that thus can be considered part of their internal quasi-electron wave functions.   Conversely, one unit of flux from 
each neighboring electron is absorbed by electron 1, incorporated
into an anti-aligned couplings with the angular momentum of quasi-electron 1. This removes another
$N-1$ quanta from electron 1, while produce a scalar that we can regard as the translationally invariant 
intrinsic wave function of quasi-electron 1, as in
Eq. (\ref{eq:dc0}).  This leaves an uncanceled angular momentum of $S-N+1=\ell-s$ in quasi-electron 1.  Thus we identify
$\ell-s$ with $S^{eff}$.  As number of occupied sub-shells is $p=2s+1$ and the number of electrons $N=(2\ell+1)(2s+1)$, one can express the
resulting $S^{eff}/S$ (the reduction in the effective flux), for large $N$, as $1/(2p+1)$.

Thus the standard result for the reduced magnetic field $B^{eff}$ is obtained in the present quasi-electron picture, while each quasi-electron
involves only a single magnetic flux spinor.\\

\subsection{Sub-shell Structure}
The space of hierarchy and conjugate
states is two-dimensional, defined by $\ell,s$.  Thus various one-dimensional
``trajectories" in the $(\ell,s)$ (or, equivalently, $N,S$) parameter space can be
followed, to illustrate the rather interesting evolution of the quasi-electron sub-shell structure.   We consider three cases:
\begin{enumerate}
\item States of fixed $\nu$, which for hierarchy states corresponds to fixed $s$, with $N$ incremented by running $\ell \ge s$
($(2\ell+1)(2s+1)=N$);  for conjugate states, $\ell$ is fixed, and $N$ is incremented by running $s \ge \ell$. 
\item Half-filled or self-conjugate states: this includes the trajectory where $\ell=s$ and thus $N=(2s+1)^2$.
\item States of fixed $S$ but running $N$, describing the successive additions of electrons while
the magnetic field $B$ is kept fixed.
\end{enumerate}

\noindent
{\it States of fixed $\nu$, for fixed $s$ or $\ell$:} 
The figures group
hierarchy states with their conjugates analogs -- defined by the exchange $(\ell,s) = (\bar{s}, \bar{\ell})$, where we add bars
to distinguish the conjugate state quantum numbers from those of their hierarchy-state analogs.
This identification connects the associated fillings according to
\[ \nu = {p \over 2p+1} =  {2 s+1 \over 4 s+3} ~ \leftrightarrow ~ {2 \bar{\ell} +1 \over 4 \bar{\ell} +1} ={p \over 2p-1} =\bar{\nu}, \]
associating $\nu=1/3$ with $\nu=1$, $\nu=2/5$ with $\nu=2/3$, etc.   When described in terms of electrons,
the paired states have no obvious common structure and reside in different Hilbert
spaces (distinct values of electron angular momenta $S$).  But in their quasi-electron representations, they have the same
sub-shell structure and occupations and a common quasi-electron angular momenta $S-N+\mathcal{I}$. 

 All states of fixed $\nu$ begin 
with an elementary state with $\ell=s$,
the configuration of minimum $N$ in which the requisite number of sub-shells is occupied: $2s+1$ in
the case of hierarchy states, $2\ell+1$ in the case of conjugate states.  Every choice $\ell,s$ is associate with
a state of definite $\nu$ with the exception of the elementary $\ell=s$ states, which are the first members
of both hierarchy and conjugate series, depending on whether $s$ or $\ell$ is kept fixed 
as $N$ is incremented.  Thus the are members of both the $\nu$ and $\bar{\nu}$ series.

\begin{figure*}
\begin{center}
\includegraphics[width=0.75\textwidth]{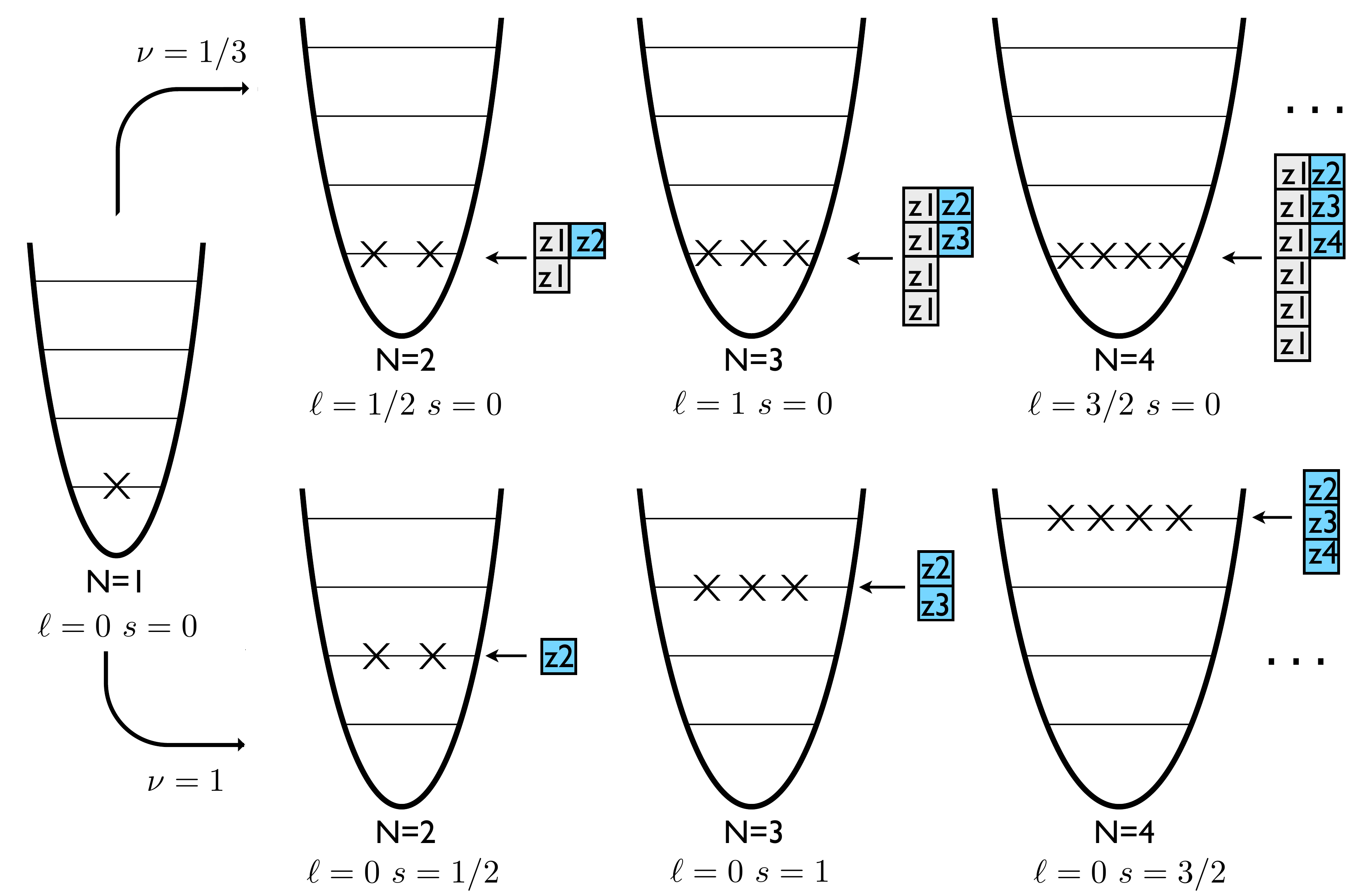}
\end{center}
\caption{Quasi-electron representations for the $\nu=1/3$ hierarchy states and $\nu=1$
conjugate states.  This Laughlin case is the simplest example of a hierarchy/conjugate pair,
corresponding to a single filled sub-shell.  There is a common ``seed," the elementary single-electron state, and
an algebraic correspondence between the shell structures of the two cases, at each electron number $N$.   
Different sub-shells are occupied, however, as the quasi-electrons for the $\nu=1/3$ and $\nu=1$  occupy 
shells with $\mathcal{I}=1$ and $N$, respectively.}
\label{fig:131}
\end{figure*}

\begin{figure*}
\begin{center}
\includegraphics[width=0.94\textwidth]{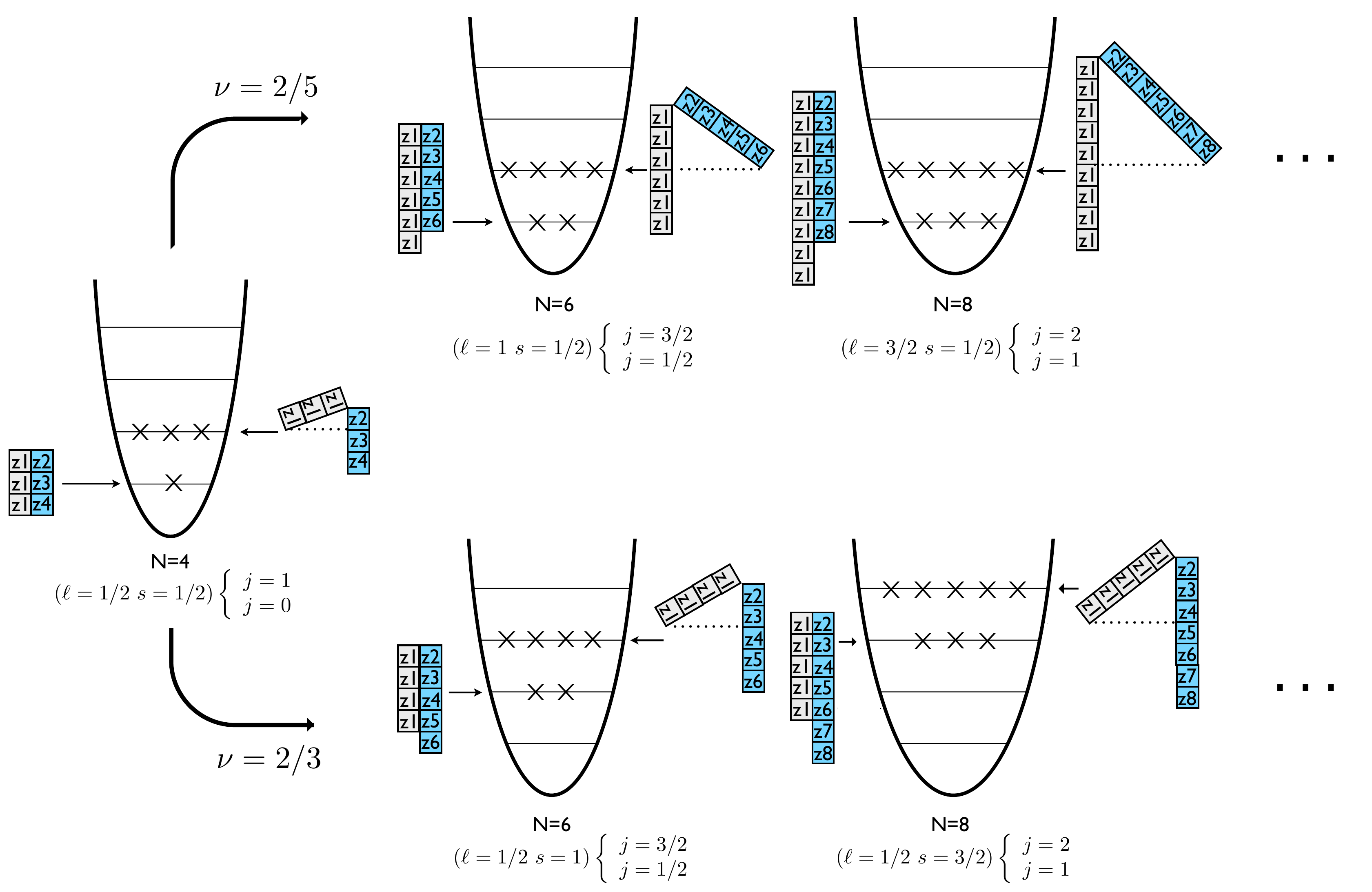}
\end{center}
\caption{As in Fig. \ref{fig:131}, put for the two-sub-shell case of the $\nu=2/5$ hierarchy and $\nu=2/3$
conjugate states.  The geometric evolution of these states is again exceedingly simple.  Sub-shells, as before,
are indexed by the number of scalars $z_1-z_i$ missing from their respective quasi-electrons.
Further discussion is given in the text.}
\label{fig:2523}
\end{figure*}


Figures \ref{fig:131} and  \ref{fig:2523} 
show  the one- and two-sub-shell cases, correspond to the fillings $1/3 \leftrightarrow 1$ and $2/5 \leftrightarrow 2/3$, respectively.
In Fig. \ref{fig:131}, as electrons are added, an algebraic
correspondence between the sub-hells is maintained: the occupied sub-shells and quasi-electrons have the same rank, at each
$N$.  The states differ because the shells are labeled by different values of $\mathcal{I}$: as electrons are added, a pairing
gap opens up between occupied hierarchy and conjugate shells.
No scalar pairs are formed in the recursion for $\nu=1$, so consequently the occupied shell is that with $\mathcal{I}=N$.
In contrast, a scalar pair accompanies the addition of each electron for $\nu=1/3$, so $\mathcal{I}=1$, independent of $N$.

Figure \ref{fig:2523} shows the two-shell cases of $\nu=2/5$ and 2/3.   This structure 
constrains $N$ to increase by two, at each step, as a quasi-electron is added to each sub-shell.
The occupied sub-shells for $\nu=2/5$ are $\mathcal{I}=1$ and 2: the $(N+1)/2$ quasi-electrons
in the upper sub-shell are missing a scalar pair with exactly one of their neighbors.   The occupied
sub-shells for $\nu=2/3$ are labeled by $\mathcal{I}=N/2-1$ and $N/2$.   Thus a pair gap again
opens up between the hierarchy and conjugates states, but at a rate in $N$ half that found
for the case illustrated in Fig. \ref{fig:131}.

The pattern of Fig. \ref{fig:2523} continue for the pairs $\nu=3/7$ and 3/5, $\nu=4//9$ and 4/7, etc., but with 3, 4, $\dots$, 
shells occupied.\\

\begin{figure*}
\begin{center}
\includegraphics[width=0.80\textwidth]{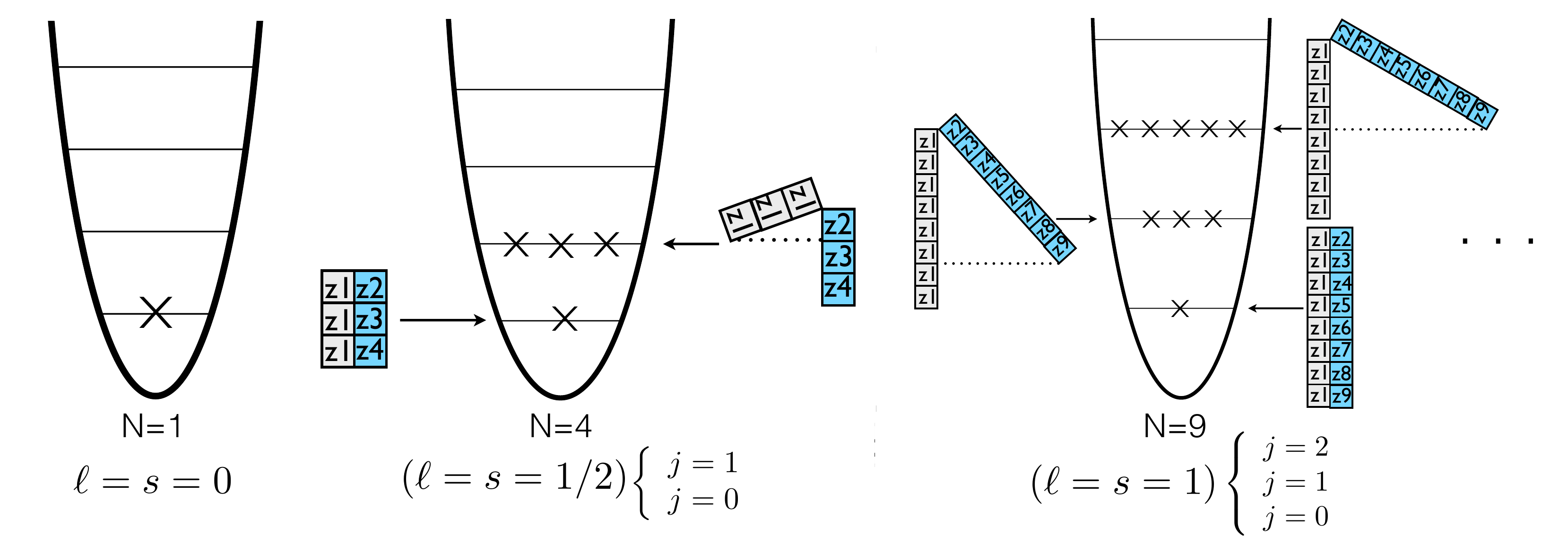}
\end{center}
\caption{The series of closed-sub-shell quasi-particles states defined by $\ell=s$, converging to the 
state with $\nu=1/2$.}
\label{fig:half}
\end{figure*}

\noindent
{\it The $\nu=1/2$ self-conjugate series: $\ell=s$:} The sub-shell structure defined by fixed $s$ (hierarchy states)
and fixed $\ell$ (conjugate states) corresponds to paths of fixed $\nu$ in the $\ell, s$ or $N, S$ plane.
Another path of fixed $\nu$, converging to the half-filled shell, is defined by
$\ell=s$ and thus $N=(2\ell+1)^2=(2 s+1)^2$=1, 4, 9, $\dots$.  This path
is confined to the self-conjugate incompressible states: when $N=1$, $\ell=s=0$, the $\nu$=1/3 and $\nu$=1 states 
coincide; when $N=4$, $\ell=s=1/2$, the $\nu=2/5$ and $\nu=2/3$ states coincide;
when $N=9$, $\ell=s=1$, the $\nu=3/7$ and $\nu=3/5$ states coincide; etc.  Thus the series can be viewed
as converging, in the large $N$ limit, to the half-filled shell simultaneously from the high- and low-density sides.
This series is illustrated in Fig. \ref{fig:half}.  

Similar series can be generated for paths defined by $\ell-s=i$, where i is a fixed half integer or integer.  All in the large
$N$ limit correspond to $\nu$=1/2.  
This observation is a restatement of the fact that hierarchy and conjugate states are
dense near $\nu=1/2$:  one can reach these states by trajectories of fixed $\nu = 1/2$ as well as by 
dialing $\nu$ from the low-density or high-density sides.

Below we discuss some of the symmetry properties of wave functions at $\nu=1/2$.\\

\noindent
{\it States for fixed $B$, variable N:} If we fix magnetic field $B$
(equivalently $2S$), $\nu$ will evolve from 0 to 1 as successive electrons are added.
In general this path encounters some of the incompressible
closed-sub-shell shell states but not all, as $\nu=2/5$ closed-shell quasi-electron states are found 
only if $N$ is even, $\nu=3/7$ only if $N$ is divisible by three, etc.  The path runs through
quasi-electron states that are not closed shells, and thus in particular produces quasi-electron representations
of states involving the addition or subtraction of one electron to the incompressible states, forming 
particle and hole states.

For such states, as the flux through the sphere or its planar analog is fixed, the electrons have a fixed
total angular momentum $S$.  The quasi-electron for the lowest (fully anti-aligned) sub-shell then must be
\[ \left[ [z_1]^{S - {N-1 \over 2}} \otimes [z_2 \cdots z_N]^{N-1 \over 2} \right]^{S-N+1} \]
as there are $N-1$ other quasi-electrons each containing one factor of $u(1)$.   Consequently every time
$N$ is incremented by a unit, the angular momentum of the lowest sub-shell is reduce by a unit, and thus the
maximum occupation of that sub-shell is reduced by two.  Other quasi-electron sub-shells evolve
similarly.  That is, the addition of an electron alters the sub-shell structure and the pattern of shell
occupations: this is not the fixed-shell structure familiar from atomic physics, for instance.

Figure \ref{fig:FixedMag} shows the evolution of states from $\nu \sim 1/3$ to $\nu=1$, as $N$ is incremented,
for $S=15$.  The choice of $S$ is clearly arbitrary, but $S=15$ supports a self-conjugate state ($N$=16,
$\nu$=4/9 and well as $\nu=4/7$) as well as other closed-sub-shell quasi-electron states ($N$=11, $\nu$=1/3; 
$N$=15, $\nu$=3/7; $N$=20, $\nu$=2/3; $N$=31, $\nu$=1).  The sub-shells should be viewed as a
procedure for defining a low-momentum Hilbert space, within which one can construct zeroth-order
wave functions, as one normally does in an effective theory.  The figure shows that the Hilbert spaces
one forms by filling the lowest sub-shells not only describe the incompressible states (those states where
the Hilbert space contracts to a single state, and where that state is a scalar) but also other fillings, where 
multiple low-energy states can be formed.  Thus one can use the quasi-electron representation to
describe low-lying states of arbitrary filling.

A number of the features of Fig. \ref{fig:FixedMag} are generic, not dependent on the specific choice 
of $S=15$.  Examples include the quasi-electron representations of particle and hole states obtained by
adding or subtracting a particle from one of the incompressible states.  The simplest cases are the
particle and hole excitations of the $\nu$=1/3 states, which one sees from the figure correspond to three-quasi-electron
particle and hole states, respectively.  As the angular momentum zero state is unique, corresponding to 
the-quasi-electron coupling $|((jj)J_{12}=j, j,J=0 \rangle$,
the quasi-electron representation of the low-momentum states predicts that there is a single 
low-energy translationally invariant, homogeneous particle (or hole) state built on the $\nu=1/3$ state.\\

\subsection{Composite Fermion and Hierarchical Constructions}
There have been recent discussions about the distinctions between or equivalence of 
hierarchical descriptions of the FQHE -- initially proposed by
Haldane \cite{Haldane} and by Halperin -- and the CF picture advocated by Jain \cite{JainCF}. Given
the explicit construction presented here, it is interesting to see how GH$^2$ description fits into the context of these two
views of the FQHE.

The GH$^2$ construction is in good accord with the CF picture of Jain, though differences are also
apparent:
\begin{enumerate}[leftmargin=0cm,itemindent=.5cm]
\item The GH$^2$ sub-shell structure is contained within the FLL, and thus the incompressible states that corresponding to the
filling of these sub-shells also involve only FLL degrees of freedom.  The angular momentum substructure derives from the fact that the
breaking of scalar pairs that must accompany increases in the density generate angular momentum.  One minimizes the Coulomb
by breaking the fewest such scalar pairs, which then identifies the fractional fillings with the nondegenerated states formed from 
completely filling the lowest sub-shells.
\item  The standard description of a CF is an electron coupled to two units of magnetic flux, while the GH$^2$ quasi-electrons
incorporate only a single magnetic flux unit into their intrinsic wave functions.  However both descriptions agree that, with the
addition of each electron, two units of magnetic flux become intrinsic (that is, combined into scalars, and thus not contributing to
net quasi-electron angular momentum).  In the 
GH$^2$ construction, only one of those flux units is carried by the added electron:
the second is divided, one quantum each, among the $N-1$ pre-existing quasi-electrons, absorbed into their internal wave functions.
\item The incompressible FQHE states are not formed by filling successive sub-shells by equivalent CFs.  Rather, for a
fixed $B$ and thus $S$, there is a maximum number of Laughlin-like quasi-electrons that can be formed in the Hilbert space.  When
that limit is reached (at $\nu=1/3$), a new ``flavor" of quasi-electrons forms, involving an additional broken pair, and carrying
one additional unit of angular momentum so that a second sub-shell arises..  This ultimately leads to $\sqrt{N}$ distinct sub-shells, with their respective quasi-electrons
co-existing in the Hilbert space at $\nu=1/2$, the most complex case.
\end{enumerate}
Though these differences from the standard CF description are of some consequence, in total the picture that emerges is
conceptually compatible with the basic ideas about CFs.  There is a mapping of the interacting electron, open-shell FQHE problem
into a closed-sub-shell noninteracting quasi-electron one.  There is a sub-shell structure -- though it is a fine-structure, unrelated to the IQHE, 
with the gap between
neighboring sub-shells associated with a single broken pair per electron.  There is a reduced magnetic
field:  the reduction explicitly arises
from the anti-alignment of the angular momentum that the electrons would carry in the absence of Coulomb interactions,
with the angular momentum generated by the successive
breaking of scalar pairs, the variational response of the electrons to minimizing their Coulomb interactions.

Aspects of the GH$^2$ construction are also hierarchical.    First, the GH operators are strictly hierarchical, forming a
tower where the next object is obtained from the previous one.  Recall that these operators create fillings
$\nu=p/(2p+1)$ by acting on the half-filled shell, both creating the necessary overdensities among $p$ electrons (relative $\nu=1/3$) that such
fillings require, then guaranteeing that these region are separated by a certain minimal number of quanta.  Denoting the
electron labels of the two clusters as $\{ 1,2,\cdots \}$ and $\{a,b, \cdots \}$, the series $p$=1,2,3,... involves the progression
\begin{equation}
\begin{array}{lrcl} \mathcal{I}=1 ~~& u_1 &\cdot& u_a \\
\mathcal{I}=2 ~~& d_1 \cdot d_2~ u_2 &\cdot& u_b ~d_a \cdot d_b \\
\mathcal {I}=3 ~~& d_1 \cdot d_3~ d_2 \cdot d_3 ~u_3 &\cdot& u_c ~d_a \cdot d_c ~d_b \cdot d_c  \\
 &  & etc. & \end{array} \nonumber
\end{equation}
where the operator at step $\mathcal{I}$ is given by the product of the factor in the $\mathcal{I}$th row above with and those from all previous steps. 
There is an obvious and simple recursion relation generating the GH operators.

However the most common description of a hierarchical structure is at the wave function level,
the conjecture that successively more elaborate CFs are 
created through Laughlin-like correlations among the CFs from previous steps
in the hierarchy.   Apart from differences already noted, the sub-shell structure of the GH construction identifies
Jain's CFs with the quasi-electrons we have defined here, all of which have a common simple structure and are not hierarchical,
at least as this term is commonly used.
But a hierarchy is an appropriate description of the over-densities that are created, which evolve
from single quasi-electrons at $\nu=1/3$ to a collection of $\sqrt{N}$ such objects at $\nu=1/2$.  For example,
consider the process of adding quasi-electrons to create a state of filling $\nu=p/(2p+1)$ with the next
highest electron number.  The process is simple -- one quasi-electron $\psi_{\mathcal{I}}$  must be added to each 
of $p$ sub-shells -- and as a series in $p$ is hierarchical,
\begin{equation}
\begin{array}{lc} p=1 ~~&\{ \psi_{\mathcal{I}=1} \}~~~~~~~~~~~~\ \\
p=2 & \{ \psi_{\mathcal{I}=2}, \psi_{\mathcal{I}=1} \}~~~~~~~~~~~~\ \\
p=3 & \{\psi_{\mathcal{I}=3}, \psi_{\mathcal{I}=2}, \psi_{\mathcal{I}=1} \}~~~~~~~~\mathrm{etc. }
   \end{array} \nonumber
\end{equation}
This process
is accompanied by the addition of magnetic flux: as the starting and new states are both scalars, the added quasi-electrons
and the added flux must combine to form a scalar.  The steps to create a new state for, say,
$p=4$, are those required to create a new state for $p=3$, except that a fourth $\mathcal{I}=4$ quasi-electron must now be 
included.

The GH$^2$ construction thus supports the general features
of both  CF and the hierarchical descriptions, and suggests that some of the debate over the CF/hierarchy issue
may derive from assuming the hierarchical objects are the CFs: in the present GH$^2$ construction,
the most natural hierarchical structures are the collections of CFs that are involved in the recursions illustrated in
the figures, and that form the regions of over or under density.

\subsection{Symmetries at $\nu=1/2$: Majorana and Pseudo-Dirac Properties of the Quasi-electrons}
The states near the half-filled shell have long raised important questions.  Halperin, Lee,
and Read argued that the ground state at $\nu=1/2$ is a Fermi liquid \cite{HLR}.  The sequences of GH$^2$ states with 
$\ell-s=$constant converge to $\nu=1/2$ in the large $N$ limit.  The special
state with $\ell=s$, illustrated in Fig. \ref{fig:half}, is defined by a vanishing $B^{eff}$: we
previously noted that the effective residual field experienced by the quasi-electrons is determined by $S^{eff}=\ell-s$.
This state can be viewed as the limit of a series of states that can be
simultaneously labeled as $\nu=p/2p+1$ and $p/2p-1$.  Although the limit yields an even-denominator 
state, the limit is reached through a series of conventional odd-denominator states.  

The states along this trajectory consist of $\sqrt{N}$ sub-shells
containing an average of $\sqrt{N}$  particles.  Thus the states at vanishing $B^{eff}$ correspond to the case where
the depth of the quasi-electron Fermi sea is maximal, scaling as $\sqrt{N}$.   In contrast, for any of the conventional fillings
characterized by fixed $s$ (hierarchy states) or $\ell$ (conjugate states), the quasi-electron Fermi seas have a finite
depth consisting of a few sub-shells, $2s+1$ or $2 \ell+1$,  in the large $N$ limit.  Because sub-shell splitting are identified 
with successive removals
of one factor of $z_1-z_i$ from electron 1, the GH$^2$ construction yields at $\nu$=1/2
\[ E_F \sim \sqrt{N} ~\alpha \left( {\hbar c \over a_0} \right) ~~~~~ A_\mathrm{defect} \sim \sqrt{N} 2 \pi a_0^2 \]
That is, the energy of the quasi-electron Fermi sea is on the order 
of $\sqrt{N}$ of the two-electron p-wave energy of Eq. (\ref{eq:Coulomb}), and as the defects involve
$\sqrt{N}$ electrons, the typical area covered by the density perturbations is on the order of $\sqrt{N}$
times the area available to each electron.\\

\noindent
{\it $\ell \leftrightarrow s$ symmetry:} There are some suggestive aspects of this interesting state at $\ell=s$ that we describe now, that support the 
notion that the physics near $\nu=1/2$ is nontrivial.  Despite being at $\nu=1/2$, the states with $\ell = s$ 
and $B^{eff}=0$ are not particle-hole symmetric, as
$S=N-1$ and thus $N/2S+1=N/(2N-1) \ne 1/2$.  Under a particle-hole transformation with respect to the electron or
electron-hole vacuums (the IQHE states at $\nu=0,1$), these $N$-electron states 
transform into $N-1$ electron states, and thus not into themselves.  However these states
have an exact symmetry connected with the interchange $\ell \leftrightarrow s$.  Let $C$ be the operator that exchanges 
single-particle spinors with Schur polynomial
spinors, that is, particle creation with magnetic flux creation
 \[                         C:  [z_1]^{N-1 \over 2} \leftrightarrow [z_2 \cdots z_N]^{N-1 \over 2} \]
Under this operation the quasi-electrons transform as
\[  C ~\Psi_{N,L,L_z,\mathcal{I}}(z) \rightarrow (-1)^{N+\mathcal{I} }\Psi_{N,L,L_z,\mathcal{I}}(z) \]
They behave like Majorana states under this operation, transforming to themselves up to a sign, with quasi-electrons 
in neighboring sub-shells having opposite C.

The pattern of neighboring sub-shell of opposite $C$ is reminiscent of particle physics scenarios in which
two (degenerate) Majorana neutrinos of opposite CP are patched together to form a Dirac neutrino, which
then transforms under CP not to itself, but to its anti-particle.  A similar transformation cannot be performed for
a fully spin-polarized system, as that would requires combining states of different angular momentum.  But in
a two-component system that might arise in the limit of small $B^\mathrm{eff}$, it would be possible.
One can make this transformation sequentially, 
 beginning with the electron in the $\mathcal{I}=1$ sub-shell, and proceeding to the top of the Fermi sea.  Denoting electron
spin by $\Sigma=1/2$ and the coupled anwe can form the states
\[ | \mathcal{I}=1; J=\tfrac{1}{2}, M \rangle_\pm \equiv \left( \begin{array}{c} ~~|\mathcal{I}=1;(L=0~ \Sigma)J={1 \over 2}, M \rangle \\  ~ \\ \pm  |\mathcal{I}=2; (L=1~ \Sigma)J={1 \over 2}, M \rangle \end{array}  \right) \]
which then transform as Dirac states,
\[ C | \mathcal{I}=1; J=\tfrac{1}{2}, M \rangle_\pm \rightarrow(-1)^{N + \mathcal{I}} | \mathcal{I}=1; J=\tfrac{1}{2}, M \rangle_\mp .\]
This first step has used the $2p_{1/2}$ part of the $\mathcal{I}=2$ sub-shell but not the $2p_{3/2}$ components, so
we now form
\[ | \mathcal{I}=2; J=\tfrac{3}{2}, M \rangle_\pm \equiv \left( \begin{array}{c} ~~|\mathcal{I}=2;(L=1~ \Sigma)J={3 \over 2}, M \rangle \\  ~ \\ \pm  |\mathcal{I}=3; (L=2~ \Sigma)J={3 \over 2}, M \rangle \end{array}  \right). \]
One can proceed in such steps, until the Fermi surface is reached.

For occupied states in a two-component system, a transformation to a new set of basis states has no
physical consequence.  Thus this transformation is of potential interest only at the Fermi surface, where for 
the $\ell=s$ $\nu=1/2$ state, it represents a transformation between occupied ( $\mathcal{I} = \sqrt{N}$) and unoccupied ( $\mathcal{I} = \sqrt{N}+1$) states.  One can
compare the ground state of the GH construction - a closed
$L \Sigma$ state where the natural basis is Majorana -- and the alternative of a closed $(L \Sigma)J$ sub-shell, where
the basis states are Dirac, in the case of a two-component system.   This second case requires the 
coupling of two Majorana states at the Fermi surface with energies, as argued above, $\sim E_F \sim \sqrt{N} ~\alpha \left( {\hbar c \over a_0} \right)$, but split by an energy corresponding to a single additional broken pair.
Thus the Dirac states at the Fermi sea are pseudo-Dirac -- the two coupled states are nearly degenerate, but
not exactly so. 
One concludes that a qualitatively different Fermi surface could arise from relatively modest
perturbations.    Possibly there are connections
to recent work by Son \cite{Son}, who noted that two conventional Jain-like hierarchy/conjugate sequences of different filling, 
$p/(2p+1)$ and $(p+1)/(2(p+1)-1)$, could be mapped into the same half-integer filling factor $(p+1/2)/(2p+1)$ of
a Dirac composite fermion.\\

\noindent
{\it Particle-hole symmetry:}  Consider any hierarchy state corresponding to some choice of $s$, so that $\nu=(2s+1)/(4s+3)$.  $N$ is
determined by $\ell \ge s$, $N=(2 \ell+1)(2s+1)$, and $S=N+\ell-s-1$.  Now consider 
a conjugate series.  For clarity we label the conjugate series
by $\bar{\ell}$ and $\bar{s} \ge \bar{\ell}. $  The filling is $\bar{\nu}=(2 \bar{\ell} +1)/(4 \bar{\ell}+1)$, $\bar{N}=(2\bar{s}+1)(2 \bar{\ell}+1)$, 
and $\bar{S}=\bar{N}+\bar{\ell}-\bar{s}-1.$
Then for any given $s,\ell$ we find the choice
\begin{equation}
\left. \begin{array}{c} \bar{\ell}=s+1/2 \\ \bar{s}=\ell-1/2 \end{array}  \right\} \Rightarrow \left\{ \begin{array}{l}  \nu + \bar{\nu} =1 \\ S = \bar{S} \\ N+ \bar{N}=2S+1 \end{array} \right.  \nonumber 
\end{equation}
Thus our construction produces an explicit conjugate state for every hierarchy state, with the requisite particle number to be the particle-hole (PH) conjugate.

Is the GH$^2$ conjugate state exactly the PH state?  Apart from a few small-$N$ cases, the answer is no, though the differences are very
small numerically.  However this shortcoming is a choice made in our construction: exact PH symmetry can be easily restored.  A given GH$^2$
state, once evaluated, can be readily written in second quantization.  Doing so for the hierarchy ($\ell,s$) and conjugate partner ($\bar{\ell},\bar{s}$) states above yields
\begin{eqnarray}
\psi_{N,S,\nu}^{GH^2} &=&\sum_i C^i_{m_1, .... m_N} B_{S,m_1}^\dagger  \cdots B_{S,m_N}^\dagger | 0 \rangle \nonumber \\
 \psi_{\bar{N},S,\bar{\nu}}^{GH^2} &=&\sum_i \bar{C}^i_{m_1, .... m_{\bar{N}}} B_{S,m_1}^\dagger  \cdots B_{S,m_{\bar{N}}}^\dagger | 0 \rangle
 \end{eqnarray}
where $|0 \rangle$ is the electron vacuum, $B^\dagger_{S,m}$ is the creation operator for the electron, $m_1>m_2 \cdots >m_N$, and
the sum over coefficients $C_i$ yields a scalar contraction of these creation operators.  (This in fact is the procedure followed in the
numerical calculations of the next section.)   Introducing the destruction operator phased to carry good angular momentum
\[ \tilde{B}_{S,m} \equiv (-1)^{S-m} B_{S,-m} \]
we can form two new scalar states by PH conjugation
\begin{eqnarray}
\psi_{{N},S,{\nu}} &\equiv& \bar{\psi}_{\bar{N},S,\bar{\nu}}^{GH^2} = \sum_i \bar{C}^i_{m_1, .... m_{\bar{N}}} \tilde{B}_{S,m_1}  \cdots \tilde{B}_{S,m_{\bar{N}}} | 1 \rangle \nonumber \\
  \psi_{\bar{N},S,\bar{\nu}} &\equiv& \bar{\psi}_{N,S,\nu}^{GH^2} =\sum_i C^i_{m_1, .... m_N} \tilde{B}_{S,m_1} \cdots \tilde{B}_{S,m_N} | 1 \rangle \nonumber
  \end{eqnarray}
where $|1\rangle$ is the filled FLL, which we can employ in a redefinition of our original states
 \begin{eqnarray}
 \psi_{N,S,\nu}^{GH^2} &\rightarrow& {1 \over 2} \left(  \psi_{N,S,\nu}^{GH^2} +\psi_{{N},S,{\nu}} \right) \nonumber \\
  \psi_{\bar{N},S,\bar{\nu}}^{GH^2}  &\rightarrow& {1 \over 2} \left(  \psi_{\bar{N},S,\bar{\nu}}^{GH^2} +  \psi_{\bar{N},S,\bar{\nu}} \right) \nonumber
  \end{eqnarray}
This procedure produces analytic wave functions with exact PH symmetry.

The series of states converging to $\nu=1/2$ with $2S+1=2N$ is defined by the trajectory $\ell=s+1/2$, and thus lies
immediately to the low-density side of the $\ell=s$ trajectory.
The $(\ell,s)$ and $(\bar{\ell},\bar{s})$ cases then coincide, unlike all other cases described above.  This trajectory runs
through the hierarchy states $N=2,\nu=1/3$ ($\ell=1/2,s=0$),   $N=6,\nu=2/5$ ($\ell=1,s=1/2$),  $N=12,\nu=3/7$ ($\ell=3/2,s=1$), $\dots$,
leading at large $N$ to a $\nu=1/2$ state distinct from that for $\ell=s$.   Unlike the cases described just above,
where the particle and hole states have $N \ne \bar{N}$, this trajectory has $N$=$\bar{N}$.  Thus there is only a
single trajectory of $\nu=1/2$, fully spin-polarized, self-conjugate states defined by
\[  {1 \over 2} \left(  \psi_{N,S,\nu=1/2,\ell=s+1/2}^{GH^2}+\psi_{N,S,\nu=1/2,\ell=s+1/2}\right) \]
leading to a single PH-symmetric state at $\nu=1/2$.   A PH-symmetric theory of the Fermi liquid ground state at $\nu=1/2$
has recently been discussed \cite{Son}.

\begin{figure*}
\begin{center}
\includegraphics[width=0.98\textwidth]{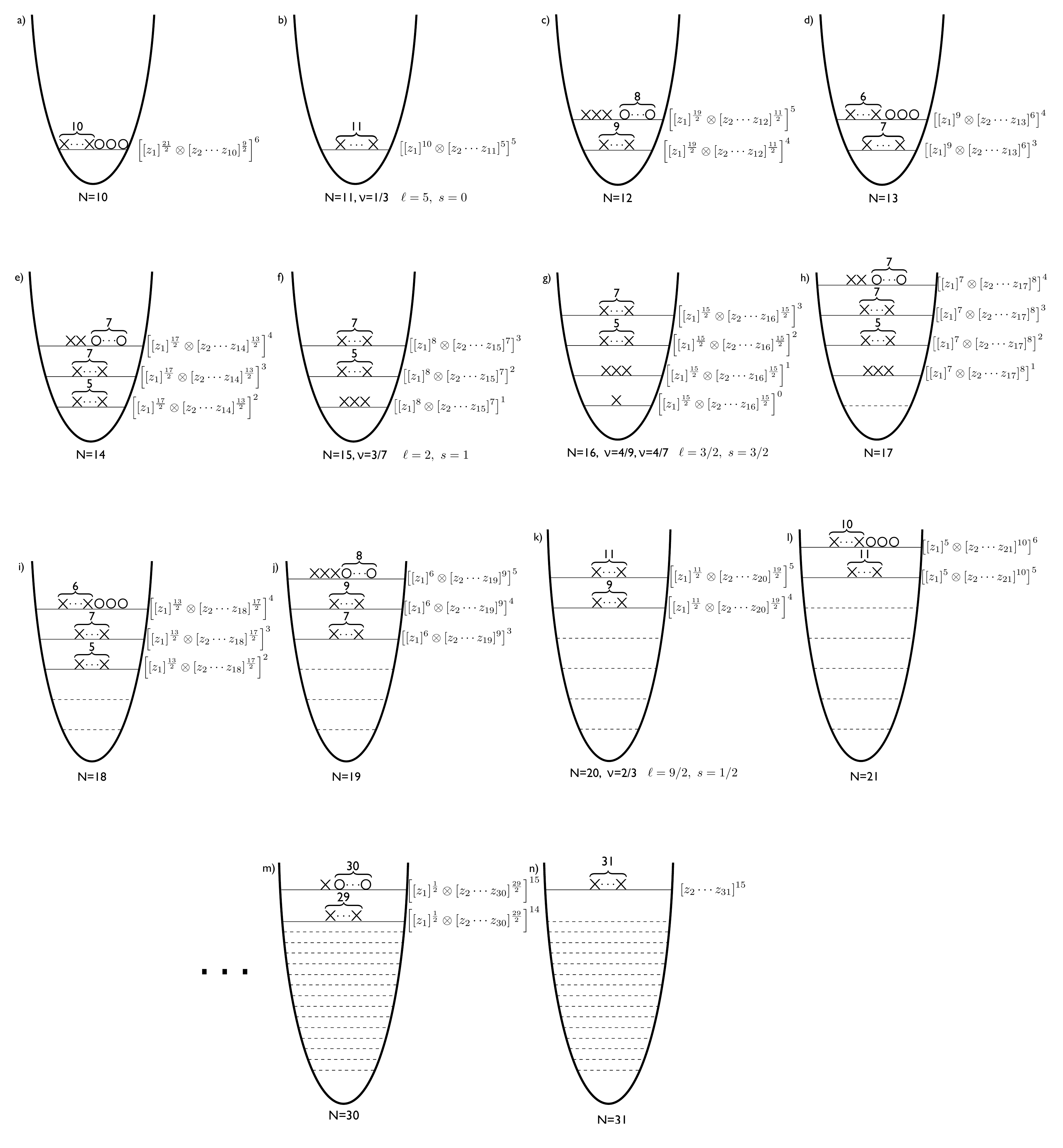}
\end{center}
\caption{The evolution of states for fixed strength of the magnetic field (fixed $S$) with increasing $N$, in terms of the
underlying quasi-electrons and their indicated sub-shells.  Here X (O) represents an occupied (unoccupied) quasi-electron state. The pattern is
illustrated for the choice $S=15$, which leads
to filled-sub-shell states with b) $\nu$=1/3 ($N=11$, $\ell=5$, $s=0$), f) $\nu$=3/7 ($N=15$, $\ell=2$, $s=1$),
g) the self-conjugate state $\nu=4/9$ and $\nu=4/7$ ($N=16$, $\ell=3/2$, $s=3/2$),  k) $\nu=2/3$ ($N=20$,
$\ell=9/2$, $s=1/2$), and n) $\nu=1$ ($N=31$, $\ell=15$, $s=0$).  Between these filling we find 
open-sub-shell quasi-electron states representing the low-energy spectrum.  The sub-shells are indexed by $\mathcal{I}=1,2, ...$,
with $\mathcal{I}=1$ the lowest sub-shell, and with $\mathcal{I}-1$ being the number of missing elementary scalars
in the indicated quasi-electron.  For $\nu>1/2$ we indicate with dash lines those sub-shells for which no quasi-electron with the requisite correlations can be constructed: sub-shells with $\mathcal{I}<N-S$ are unoccupied.
We omit diagrams for electron numbers 22-29, as they correspond to an obvious interpolation
between panels l) and m).  In panels m) and n) sub-shell spacings have been compressed in order to keep these figures compact.}
\label{fig:FixedMag}
\end{figure*}

\noindent
\subsection{The GH$^2$ States as Fermion Excitations of the Half-filled Shell: Pfaffian-like States}
This GH$^2$ quasi-electrons and wave functions can be written as fermion operators 
acting on the bosonic $\nu=1/2$ state.  Using
Eqs. (\ref{eq:dc0},\ref{eq:dc1},\ref{eq:dc2}), we find
\begin{eqnarray}
 \Gamma^{N,\mathcal{I}=1}_{L,m}(i) &=&  [z_i]^{L}_m   \nonumber \\
 \Gamma^{N,\mathcal{I}=2}_{L,m}(i) &=&  \sum_{j \ne i} {[[z_i]^{\ell-s+(\mathcal{I}-1)/2} [z_j]^{(\mathcal{I}-1)/2}]^{L}_m \over z_i-z_j}  \nonumber \\
 \Gamma^{N,\mathcal{I}=3}_{L,m}(i) &=& \sum_{ j<k;~ j,k \ne i} {[ [z_i]^{\ell-s+(\mathcal{I}-1)/2} [z_j z_k]^{(\mathcal{I}-1)/2}]^{L}_m  \over (z_i-z_j) (z_i - z_k)} \nonumber \\
 \Gamma^{N,\mathcal{I}=4}_{L,m}(i) &=& \sum_{ j<k<l;~ j,k,l \ne i} { [[z_i]^{\ell-s+(\mathcal{I}-1)/2} [z_j z_k z_l]^{(\mathcal{I}-1)/2}]^{L}_m \over (z_i-z_j) (z_i - z_k)(z_i-z_l)}   \nonumber 
 \end{eqnarray}
 and so on, where $L=S-N+\mathcal{I}=\ell-s+\mathcal{I}-1$.  The numerators are all aligned couplings.  The quasi-electrons can be written
 \begin{equation}
 \Psi_{N,L,m,\mathcal{I}}(z_i) = \Gamma^{N,\mathcal{I}}_{L,m}(i) ~R_N(i) 
 \end{equation}
 
In the case of the $\nu=1/2$ state reached through the series $\ell-s=0$, the above results can be rewritten as
\begin{eqnarray}
 \Gamma^{N,\mathcal{I}=1}_{L=0,m}(i) &=&  1  \nonumber \\
 \Gamma^{N,\mathcal{I}=2}_{L=1,m}(i) &=&  \sum_{j \ne i} {[z_i z_j]^1_m \over z_i-z_j}  \nonumber \\
 \Gamma^{N,\mathcal{I}=3}_{L=2,m}(i) &=& \sum_{ j<k;~ j,k \ne i} {[[z_i z_j]^1 \otimes [z_i z_k]^1 ]^{2}_m  \over (z_i-z_j) (z_i - z_k)} \nonumber \\
 \Gamma^{N,\mathcal{I}=4}_{L=3,m}(i) &=& \sum_{ j<k<l;~ j,k,l \ne i} { [ [z_i z_j]^1 \otimes [z_i z_k]^1 \otimes [z_i z_l]^1]^3 \over (z_i-z_j) (z_i - z_k)(z_i-z_l)}   \nonumber 
 \end{eqnarray}
 These have the form of particle-hole operators on the bosonic $\nu=1/2$ state, with
 the number of particle-hole excitations $\mathcal{I}-1$.  $R_N(i)$ is a product of $N-1$
 scalar pairs.  The denominators destroy $\mathcal{I}-1$ of these scalar pairs, defining the
 holes, while the numerators create their replacements, the corresponding aligned 
 pairs $[z_i z_j]^1$, the particles.  $R_N(i)$, as a filled ``sea" of scalar pairs,  plays the role of the 
 particle-hole vacuum.
 
Related operators have been discussed previously.  The Moore-Read Pfaffian $\nu=1/2$ state,
most often discussed in connection with the $\nu=5/2$ state, takes the form
\begin{eqnarray}
\psi(z_1, \cdots, z_{2N}) = \mathrm{Pf} \left( {1 \over z_i-z_j} \right) \prod_{i<j} (z_i-z_j)^2 ~~~~\nonumber \\
\mathrm{Pf} \left(A_{ij} \right) \equiv \epsilon_{i_1 i_2 \cdots i_{2N}} ~A_{i_1i_2} A_{i_3 i_4} \cdots A_{i_{2N-1} i_{2N}} ~~~~
\label{eq:Pf}
\end{eqnarray}
The Pfaffian operator has the same filling
as the GH$^2$ operator with $s=\ell+1/2$, the first series on the conjugate side of $\ell=s$, just as the PH-symmetric
$\nu=1/2$ series is the first on the hierarchy side.  

The $s=\ell+1/2$ GH$^2$ operator is formed from the antisymmetrized product of the single quasi-electron operators
\begin{eqnarray}
 \Gamma^{\mathcal{I}=2}(i) &=&  \sum_{j \ne i} {[z_j]^{1/2} \over z_i-z_j}  \nonumber \\
 \Gamma^{\mathcal{I}=3}(i) &=& \sum_{ j<k;~ j,k \ne i} { [z_i z_j z_k]^{3/2}  \over (z_i-z_j) (z_i - z_k)} \nonumber \\
 \Gamma^{\mathcal{I}=4}(i) &=& \sum_{ j<k<l;~ j,k,l \ne i} { [z_i^2 z_j z_k z_l]^{5/2}  \over (z_i-z_j) (z_i - z_k)(z_i-z_l)} \nonumber
 \end{eqnarray}
The GH$^2$ operator analogous to the Pfaffian is a simple determinant, e.g.,
taking the form for $N=12$
\[ \Phi^{\mathrm{GH}^2}_{N=12,s=3/2,\ell=1}  \equiv  \left| \begin{array}{ccc} \Gamma^{\mathcal{I}=4}(1)& \cdots&  \Gamma^{\mathcal{I}=4}({12}) \\
\Gamma^{\mathcal{I}=3}(1)& \cdots&  \Gamma^{\mathcal{I}=3}({12}) \\
\Gamma^{\mathcal{I}=2}(1)& \cdots&  \Gamma^{\mathcal{I}=2}({12}) \end{array} \right| \]
The wave function generated by the Pfaffian at $\nu=1/2$ has a poor overlap with that generated by the analogous GH$^2$
operator, e.g., $\sim$ 0.87 for $N=6$.   The corresponding overlap of the GH$^2$ wave function with that computed by
diagonalizing the Coulomb interaction is 0.993.  The spin-paired Pfaffian overlaps with numerically generated wave functions for $\nu=5/2$ also appear to be rather
poor, ranging from 0.69-0.87 for $N=8-16$, according to Scarola, Jain, and Rezayi \cite{SJR}.   However the agreement
can be improved significantly, if the p-wave contribution to the potential is dialed away from its Coulomb value \cite{Morf}.

The standard Pfaffian builds in $N/2$ two-electron correlations similar to those contained in $\Gamma^{\mathcal{I}=2}$, which generates only
a subset of the quasi-electrons contributing at $\nu=1/2$.  Read and
Rezayi \cite{RR2} generalized this construction to include more complicated correlations that would allow the symmetric part of the wave function
to remain nonvanishing when $\mathcal{I}$ electrons are placed at one point.  Thus this extension has some common features with
$\Gamma^\mathcal{I}$, $\mathcal{I}>2$.  However, as we have discussed previously, optimal approximate wave functions are unlikely to arise from
constructions that consider only short-range behavior, at least in cases where systems are spin polarized and the electrons 
restricted to fill the lowest LLs first.  The GH$^2$ construction is guided by the scale invariance of the potential, and thus effectively
produces wave functions that weight in an appropriate way Coulomb contributions from all partial waves.\\

\begin{table}[h]
\caption{Overlaps of the GH$^2$ quasi-electron hierarchy wave functions with the exact wave functions computed by
direct diagonalization of the Coulomb interaction.  The corresponding results for the GH/Jain wave functions are
also given.  All results are for the principal hierarchy ($m$=3).  The GH/Jain wave function overlaps are those
available from the original GH paper.}
\begin{tabular}{ccccccc}
\hline \hline
$N$ ~&~ $S$ ~ &~ $\ell$ ~&~ $s$ ~&~ $\nu$ ~ & $|\langle \psi_\mathrm{ex} | \psi^{\mathrm{GH}^2} \rangle|$ ~& ~$| \langle \psi_\mathrm{ex} | \psi^{\mathrm{GH/Jain}} \rangle |$ \\
\hline
3 & 3 & 1 & 0 & 1/3 & 1.0 & 1.0 \\
4 & ${9 \over 2}$ & ${3 \over 2}$ & 0 &  & 0.9980 & 0.9980 \\
5 & 6 & 2 & 0 & & 0.9991 & 0.9991 \\
6 & ${15 \over 2}$ & ${5 \over 2}$ & 0 & & 0.9965 & 0.9965 \\
7 & 9 & 3 & 0 & & 0.9964 & 0.9964 \\
8 & ${21 \over 2}$ & ${7 \over 2}$ & 0 & & 0.9954 & 0.9954 \\
9 & 12 & 4 & 0 & & 0.9941 & 0.9941 \\
10 & ${27 \over 2}$ & ${9 \over 2}$ & 0 & & 09930  & -- \\
4 & 3 & ${1 \over 2}$ & ${1 \over 2}$ & 2/5 & 1.0 &   1.0 \\
6 & ${11 \over 2}$ & 1 & ${1 \over 2}$ & & 0.9997 & 0.9998 \\
8 & 8 & ${3 \over 2}$ & ${1 \over 2}$ & & 0.9994 & 0.9996 \\
10 & ${21 \over 2}$ & 2 & ${1 \over 2}$ &  & 0.9978 & 0.9980 \\
9 & 8 & 1 & 1 & 3/7 & 0.9986 & 0.9994 \\
\hline
\end{tabular}
\label{tab:one}
\end{table}

\begin{table}[h]
\caption{Overlaps of the GH$^2$ quasi-electron conjugate wave functions with the exact wave functions computed by
direct diagonalization of the Coulomb interaction.  The corresponding results for the GH wave functions are
taken from the original paper.  All results are for the principal hierarchy ($m$=3).}
\begin{tabular}{ccccccc}
\hline \hline
$N$ ~&~ $S$ ~ &~ $\ell$ ~&~ $s$ ~&~ $\nu$ ~ & $|\langle \psi_\mathrm{ex} | \psi^{\mathrm{GH}^2} \rangle|$ ~& ~$| \langle \psi_\mathrm{ex} | \psi^{\mathrm{GH}} \rangle |$ \\
\hline
4 & 3 & ${1 \over 2}$ & ${1 \over 2}$ & 2/3 & 1.0 & 1.0 \\
6 & ${9 \over 2}$ & ${1 \over 2}$ & 1 &  & 0.9929 & 0.9965 \\
8 & 6 & ${1 \over 2}$ & ${3 \over 2}$ & & 0.9939 & 0.9982 \\
10 & ${15 \over 2}$ & 2 & 0 & & 0.9873 & 0.9940 \\
12 & 9 & ${1 \over 2}$ & ${5 \over 2}$ & & 0.9840 & -- \\
9 & 8 & 1 & 1 & 3/5 & 0.9986 & 0.9994 \\
\hline
\end{tabular}
\label{tab:two}
\end{table}

\subsection{Numerical Comparisons with Exact Diagonalizations}  
The GH$^2$ quasi-electron wave functions were obtain
by applying the GH operators in an alternative way -- but a way that retains all of the symmetries of the original
construction, including translational invariance, homogeneity, and what we have argued is the best
quantum mechanical approximation to the scale-independence of the Coulomb potential.  Are the resulting 
numerical results comparable? To test this we used an m-scheme Lanczos code to directly diagonalize the
Coulomb potential on the sphere, then generated the corresponding quasi-particle approximate wave function
with a Mathematica script, evaluating the overlaps.  This was done for various states ranging up to 10 electrons,
including Laughlin's $m=3$ states, which of course are identical in the GH and GH$^2$ constructions.
The analytic wave functions were generated in the plane,
which allowed use of Mathematica's polynomial capabilities, then mapped onto the sphere, via the homomorphism
we have already described.  The comparison was done on the sphere as this was the geometry
originally used by GH.  

From Table \ref{tab:one} one can see that the original GH wave
functions (an analytic version of those Jain constructed) produce marginally better
overlaps -- but in both cases the overlaps generated are typically as close to unity as those obtained
in the $\nu=1/3$ Laughlin case.   GH also generated the higher-density conjugate states, and in this case 
differences between the GH and GH$^2$ wave functions are somewhat larger -- typically comparable to the differences
between the GH wave functions and the numerically generated exact wave functions.

The advantages of the GH$^2$ wave function are its exceptional simplicity and its explicit
quasi-electron or CF form -- the ability to express
the wave function as a single quasi-electron Slater determinant, for both hierarchy and conjugate states, and in both planar
and spherical geometry.  Is this important?  In our view, some discussions about approximate 
FQHE wave functions in the literature are misguided, treating the approximate wave function as a
representation of the true wave function.  We do not think this is a sensible viewpoint.  First, it is clear that 
the pursuit of improved wave functions is futile, as all constructions deal with some low-momentum 
portion of the Hilbert space, and thus the resulting wave function always can be improved by mixing in any 
component not in that
Hilbert space: this is the variational principle. Consequently the overlap of any approximate wave function
with the true wave function will deteriorate with increasing $N$: Because the approximate 
wave function resides in a limited low-momentum Hilbert space, in any defined area of the plane, it will
omit some high-momentum components.  If one now doubles the area, and then doubles again, the chances
that a high-momentum component will be found somewhere in the extended regions grows combinatorially.
Consequently the overlap is eventually driven to zero.  This trend can be seen in numerical results generically.

A more sensible definition
of the approximate wave function is as a projection of the true wave function to some low-momentum space -- an effective
wave function.  Because the wave functions discussed here are filled Slater determinants, the prescription for
their construction defines a projection operator $P$ onto a low-momentum Hilbert space that
contains only a single state.   
Normally in an effective theory (ET), a LO projected wave function is considered a good starting point if it
has a strong overlap with $P|\psi \rangle$, the exact solution projected onto the chosen low-momentum
Hilbert space.  By this standard the GH$^2$ wave function is a trivially exact LO wave function, as the projection $P$ only
contains one state.   An interesting question -- as the Coulomb interaction provides no scale for use as
an expansion parameter -- is the construction of corrections, to produce an improved NLO wave function.

Because the GH$^2$ wave function is based on a simple set of quasi-electron degrees of freedom, one has
a starting point that could conceivably could allow NLO corrections.  A first step in such a process is suggested
by, for example, cases similar to panels a) and c) of Fig. \ref{fig:FixedMag}, ``open-shell" quasi-electron states.
Unlike the case of the GH or Jain wave functions, the
GH$^2$ construction provides a simple but nontrivial $P$ for such cases, consisting of a set of degenerate quasi-electron
configurations.  We return to this topic below.

 \section{Summary and Future Directions}
 In this paper we have constructed an explicit quasi-electron representation of the hierarchy and conjugate
 FQHE states.  The quasi-electrons have a generic form, a vector product of an spinor that creates
 single-electron states and one formed from Schur polynomials that adds a unit of magnetic flux to all
 existing electrons.  The quasi-electrons and the sub-shell structure they induced within the FLL are
 quite novel, with both the sub-shell structure and quasi-electrons evolving as particles or magnetic flux 
 are added.  Effectively the construction explicitly maps the problem of electrons strongly interacting
 in a partially filled sub-shell, into a noninteracting problem, a single Slater determinant of quasi-electrons.
 The quasi-electrons are fermions that carry good $L$ and $L_z$, and the scalar many-body states
 they form as Slater determinants are translationally invariant and homogeneous (uniform one-body
 density).  The construction is done both on the plane and on the sphere.   The hierarchy and conjugate
 states corresponds to those fillings $\nu$ where the quasi-electron representations of the wave functions
 are unique, consistent of a set of completely filled shells.  The resulting wave function defines a low-momentum Hilbert space consisting
 of one Slater determinant that efficiently captures much of the long wave-length behavior of the FQHE.
 
 The quasi-electron SU(2) fine structure that exists in the FLL is governed by energy gaps that represent the cost of replacing a
 favorable scalar pair $[z_1 z_i]^0$ by an unfavored pair $[z_1 z_i]^1$.  As implemented
 in the GH$^2$ wave function, this replacement affects correlations at all distance scales, and thus is 
 a variational ansatz consistent with the scale-invariance of the Coulomb potential.   (Laughlin's construction
 has the same property, though it is often misconstrued as a strategy for limiting unfavorable short-range correlations.)
 The connection between correlations and
 angular momentum is natural in the FQHE, as the generator of rotations is the sum over pair-breaking
 single-electron operators.  These observations address the issues that  troubled Dyakonov and others
 about the use of multiply-occupied LLs in the Jain construction:  Jain borrowed the needed SU(2)
 algebra from the IQHE, while in fact the physically relevant SU(2) algebra comes from correlations
 within the FLL.

The construction was extended to the plane in two steps.  We first defined a truncated Hilbert space in
the plane that contains the same number of single-electron degrees of freedom as on the sphere.  Effectively
this defines a planar analog of the angular momentum operator $L$.  We then packaged these degrees 
of freedom in single-particle (electron) and Schur-polynomial (magnetic flux) spinors, so that we could 
exploit the natural mapping between $L_x,L_y$ and the $P_x,P_y$, and thus between
rotational invariance on the sphere and translational invariance in the plane.  We introduced
scalar and tensor products on the plane that preserved the homomorphism between the respective
lowering operators.  Consequently, we were able to construct in a simple procedure $N$-electron scalar
wave functions in the plane that are translationally invariant and that have a uniform one-body density
(homogeneous), where the latter property is defined by the homomorphism with the sphere.
 
 While the CF picture of the quantum Hall effect has been an important theme of the
 field for many years, we are not aware of any detailed, explicit construction of the quasi-electrons and their
 sub-shell structure within the FLL, on the plane and on the sphere, and including the full set of both
 hierarchy and conjugate states.  The quasi-electrons that emerge are qualitatively in good accord
 with the ideas of Jain, exhibiting for example the screening of the magnetic flux through the 
 absorption of magnetic flux quanta in the intrinsic (scalar) wave functions of the quasi-electrons:
 screening is manifested in anti-aligned couplings of electrons to magnetic flux.  However, there
 are differences in detail, including the form of the quasi-electrons (the coupling of an electron spinor
 to a single magnetic flux spinor, not two) and the existence of an explicit FLL fine structure, reflecting the
 fact that there are multiple types of quasi-electrons, sharing a common functional form but differing
 in their electron/magnetic flux angular momentum couplings.
 
 There are also hierarchical aspects of the construction, though the notion of hierarchical structures
 seems most relevant to the regions of over-density or under-density, or
 equivalently, to the set of $p$ distinct quasi-electrons that must be added, one to each occupied
 quasi-electron shell, to create new wave functions of the same $\nu$ but with $p$ additional electrons
 and $2p+1$ new flux quanta.  There is a clear hierarchical structure among the $p$ added quasi-electrons.
 
 Consistent with the observation that $\nu=1/2$ is the convergence point for the series $p/(2p+1)$ and
 $2p/(2p-1)$, a dense collection of GH$^2$ states can be defined as the large-$N$ limit of the 
 series $\ell-s$=constant.  Each corresponds to a deep Fermi sea of quasi-electrons, containing
 $\sim \sqrt{N}$ occupied shells.   We discussed particular symmetries of these states, including
 the $\ell=s$ trajectory that is symmetric under interchange of electron and magnetic flux spinors,
 and associated with an additive charge, and the trajectory $\ell=s+1/2$, which is the particle-hole
 symmetric case.   
 
 Finally, motivated by the Pfaffian, we described the GH$^2$ wave functions in perhaps their simplest
 form, as a operator involving only the coordinates $z_i$ acting on the bosonic half-filled shell.  The
 quasi-electrons are then generated by simple operators $\Gamma^\mathcal{I}(i)$ than act on the
 scalars $R_N(i)$, which can be regarded as the corresponding bosonic quasi-electron.  In this
 language, the $\nu=1/2$ $\ell=s$ state becomes a product of particle-hole excitations of the bosonic
 vacuum, where the holes correspond to destroyed scalars $[z_1z_i]^0$, and the particles the $[z_1z_i]^1$.
 This form of the GH$^2$ construction provides a series of operators that can be used as Pfaffians
 are typically used.  We argue that they are better motivated physically, as they do not focus
 exclusively on the short-range properties of the Coulomb potential.
 
 One of the advantages of a simple representation of wave functions is that it provides an intuitive starting
 point for understanding the physical mechanisms behind less well understood aspects of the FQHE.
 We mention two possibilities for extending the GH$^2$ construction, to make further progress.
 
 A set of states of fractional filling  3/8, 4/11, 6/17, ..., residing between $\nu$=1/3 and 1/5, have been 
discovered experimentally \cite{Pan}.  These states nominally correspond to ``open sub-shell" quasi-particle configurations
where the sub-shell that is filled at $\nu=2/5$ is instead partially filled, and thus where a set of degenerate
many-body quasi-electron states exist (see Fig \ref{fig:FixedMag}).   That is, the Hilbert space $P$ defined
by the quasi-electron construction is no longer trivial, but contains multiple states.  The situation is 
precisely analogous to the original problem that the GH$^2$ construction addresses, finding the best
simple low-momentum representation of interacting electrons that occupy an open shell:  quasi-electrons
now replace the electrons.

In the context of a nonrelativistic ET, this problem likely presents the first step toward
generalization.  Any truncation in a Hilbert space, such as that implicit in the quasi-electron construction,
must be accompanied by the introduction of an effective interaction, to account for the degrees of freedom
that have been integrated out.  That effective interaction can be derived from the full Hamiltonian; alternatively, it can be
parameterized, by fitting the associated low-energy constants to data.  We have not needed to address
this problem in our construction, because the hierarchy and conjugate states are the unique cases where
the projection $P$ is onto a single state.  In such cases the effective interaction clearly has no role
in altering the wave function: it only contributes to ground state energy shift, which we have not addressed
in this paper.  However, the case of an open quasi-electron sub-shell is different, as the LO Hilbert space defined by $P$ is
no longer trivial, and thus $H^\mathrm{eff}$ is important, as it breaks the degeneracy within $P$ and
thus determines the nature of the ground state.  Progress we have made on this problem, including
the analytic wave function that is obtained, will be described elsewhere \cite{DW}.

A second open question concerns the nature of the half-filled shell, specifically the observation that at
$\nu=5/2$ there is a strong FQHE state that may be fully spin polarized, while there is no evidence
for such a state at $\nu=1/2$  \cite{Willet,Pan2,Eisenstein}.  Descriptions of the $\nu=5/2$ state in terms of Pfaffians or anti-Pfaffians
have claimed some success.  But in the context of the present construction, the differences between
the $\nu=1/2$ and $\nu=5/2$ states presents a puzzle, as non-degenerate LO candidate states could be
constructed for either.  The LO wave function would fully populate the lowest two LLs in the case of
$\nu=5/2$, generating a contribution to the effective interaction for the valence electrons, but such a
shift in single electron energies should not affected the correlations.  This leads to the speculation
that the differences between these two states must arise from corrections to the present construction,
that is, from next-to-leading-order (NLO) contributions.

Many ETs have an obvious expansion, a small parameter such as the ratio of Yukawa masses in a potential
with both short- and long-range components.  But the Coulomb interaction contains no such scales.
Instead, the natural ET expansion for the Coulomb interaction is that used in classical electromagnetism,
the multipole expansion.  The expansion parameter is not intrinsic to the potential, but rather to the geometry:
$(R/r)^l$ where $l$ is the multipole order, $R$ is the size of the regions of charge, and $r$ their separation.
Because the FQHE is modeled as a one-component system in a neutralizing but static background,
one anticipates that the NLO corrections would be associated with $l=2$.  While in the FLL electrons
are restricted to the lowest circular cyclotron orbit, this is not the case for the $\nu=5/2$ state, where the intrinsic
cyclotron motion carries two quanta.  The orbit can respond to higher multipoles.

This problem has analogs elsewhere in physics, where the symmetry-violating effects connected with
quadrupole interactions drive objects to deform, with symmetry then restored by the associated Goldstone mode,
the correlated low-frequency collective motion of the deformed objects.  This physics is frequently most dramatic at half-filled shells,
for the reasons apparent in the FQHE: the regions of over-density or under-density involve the correlations
among the maximum number of quasi-electrons $\sim \sqrt{N}$.  That is, $(R/r)^l$ is at its
most favorable point.  The single-particle 
cyclotron orbits for valence electrons in the $\nu=5/2$ state can respond to the
interactions beyond the monopole, and presumably do so in the process of minimizing their energy variationally.
We are intrigued by the fact that the candidate ET is the multipole
expansion, that higher LLs could respond to terms beyond the monopole, and that the consequence
of such corrections are often startling, as they lead to symmetry breaking and restoration through long-wavelength
collective modes.  These possibilities should be explored.\\
   
\begin{acknowledgments}
\noindent
{\bf Acknowledgments:}  We thank Joe Ginocchio for the very enjoyable collaboration that produced the 1996
paper, Susanne Viefers for suggesting that we supply the missing Reference 12 in GH, and 
Byungmin Kang and Ken McElvain for help and comments.
This work was supported in part by the US DOE under DE-AC02-05CH11231 (LBL)
DE-SC00046548 (Berkeley),  and DE-FG02-94ER14413 (Colorado).
\end{acknowledgments}

\section{Bbiliography}

\pagebreak
\section{Appendix}
Issues differentiating the GH$^2$ construction, which was guided by the scale
invariance of the Coulomb potential, from those emphasizing control of short-range physics, as the
Coulomb potential is strongest in the p-wave, can be explored by studying the three- and four-electron problems.
 The Hilbert spaces can be defined algebraically for arbitrary $S$ in these cases.
 This appendix derives the most general four-electron wave function
 and shows its relevance to such correlation questions.  The exercise also helps to illustrate that correlations beyond the
 two-electron ones considered by Laughlin are more complicated, because they may not have simple embeddings
 in many-electron wave functions.
 
 The solution of the four-electron problem provided part of the
 motivation for the GH construction: the GH work is summarized in \cite{GHsym}.  Here we extend that
 discussion to the plane, where the basis states can be enumerated by finding all translationally invariant symmetric
 polynomials.
 
Schur's lemma tells us that the most general few-electron correlation is the product of the antisymmetric closed shell
wave function with a symmetric polynomial.  Thus it is sufficient to identify all such symmetric and
translationally invariant polynomials.
In the case Laughlin studied, two interacting electrons, those polynomials are $(z_1-z_2)^{m-1} \sim [z_1]^{(m-1)/2} \odot
[z_2]^{(m-1)/2}$, $m=1,3,5, \dots$.  The correlations are scalars, and thus also correspond
to two-electron wave functions, making their identification as well as the embedding of this correlation
in many-electron systems particularly simple:  Laughlin simply enforced this correlation on all electron
pairs.  As he did this for each pair in the most efficient way, the many-body states he constructed are those of maximum
density for which the two-electron correlation constraint can be preserved for all pairs. 

The GH construction began with the observation that other hierarchy wave functions could not be
described simply in term of two-electron correlations, as Laughlin had exhausted all such possibilities.  Furthermore
Laughlin's restriction forbidding p-wave two-electron correlations clearly cannot be maintain beyond $\nu>1/3$.
It is then natural to turn to the three-electron correlation to limit unfavorable interactions: a three-electron
constraint can guarantee that electrons correlated in a p-wave (IQHE two-electron droplets)
remain separated from any similar overdensity, minimizing the energy, at least until
some critical density is reached.  Such separation can in fact be maintained (in a defined way) until $\nu=2/5$.  

The elementary symmetric polynomials $\{1,~ f_1[z_1,z_2,z_3],~f_2[z_1,z_2,z_3],~f_3[z_1,z_2,z_3] \} =\{1,~ z_1+z_2+z_3,~ z_1 z_2+z_1 z_3 + z_2 z_3,~ z_1 z_2 z_3 \}$ form a basis for the most general symmetric
three-electron wave function:  as interaction
energies only depend on relative coordinates, we are only interested in the subset of polynomials that are
translationally invariant.  Two translationally invariant basis states can be defined, $f_2^\prime$ and
$f_3^\prime$, leading to the general form for antisymmetric translationally invariant polynomials
\begin{eqnarray}
  L[N=3,m]~ f_2^{\prime~p_2}~ f_3^{\prime~p_3}~~~~~~~~~~~~~~~~~~~ \nonumber \\
   m \in \{ \mathrm{odd~integers} \}~~~~~p_2 \in \{0,1,2 \}~~~~~p_3 \in \{\mathrm{integers} \} \nonumber \\
   f_2^\prime = f_2-f_1^2/3~~~f_3^\prime = f_3 - f_1 f_2/3 + 2 f_1^3/27~~~~~~~\nonumber \\
   f_2^\prime[z_1,z_2,z_3] \sim \mathcal{S} \left[(z_1-z_2)^2 \right]~~~~~~~~~~~~~~~~~\nonumber \\
   ~f_3^\prime[z_1,z_2,z_3] \sim \mathcal{S} \left[ (z_1-z_2)(z_1-z_3)^2 \right]~~~~~~~~~~~
   \label{eq:3b}
   \end{eqnarray}
While $f_2^\prime$ and $f_3^\prime$ are symmetric, we have broken the symmetry, then introduced the
symmetrization operator $\mathcal{S}$, to show the underlying two-electron correlations.
The restriction on $p_2$ comes from the observation that $(z_1-z_2)(z_1-z_3)(z_2-z_3))^2 \sim
4/27 f_2^{\prime~3} + f_3^{\prime~2}$, so that powers $f_2^{\prime~3}$ can be eliminated (while incrementing
$m$).  The structure is that of band-heads labeled by $m$:  For fixed m, any pair $(p_2,p_3)$ corresponds
to a unique number of quanta,  0,2,3,4,5,....   Thus degeneracies arise only when the $m=3$ band begins (9 quanta) --
that is, with Laughlin's $m=3$ state.  Thus the correlations corresponding to the higher
densities of present interest ($\nu>1/3$) are unique.

The elementary translationally invariant polynomials $f_2^\prime$ and $f_3^\prime$, however, are not
scalars and thus are not allowed states.  Specifically,
\begin{eqnarray}
f_2^\prime &\sim& \left[ [z_1 z_2 z_3]^{3/2} \otimes [z_1 z_2 z_3]^{3/2} \right]^1_{-1} \nonumber \\
f_3^\prime &\sim& \left[ [[z_1]^{1/2}\otimes [z_2 z_3]^1]^{1/2} \otimes [[z_2]^{1/2} \otimes [ z_1 z_3]^1]^{1/2} \otimes [[z_3]^{1/2}  \right. \nonumber \\
&&~~~\left. \otimes [z_1 z_2]^1]^{1/2} \right]^{3/2}_{-3/2}. 
\end{eqnarray}
where, in the expression for $f_3^\prime$, an aligned coupling of three spin-1/2 factors has been formed,
with $ [[z_1]^{1/2}\otimes [z_2 z_3]^1]^{1/2}_{-1/2} \sim  z_2 (z_1-z_3)+z_3(z_1-z_2)$.
These polynomials are translationally invariant because they are the lowest components of rank 1 and 3/2 vectors,
not because they are scalars.  They do not correspond to homogeneous wave functions.
We suspect this result has been implicitly encountered before and may have generated confusion:
unlike Laughlin's case, one cannot identify these correlations by studying three-electron
states, nor can one build scalar many-electron wave functions by simply cobbling together products of three-electron 
scalar building blocks.   It is easily proven that the only scalar three-electron states are those identified by
Laughlin, $((z_1-z_2)(z_1-z_3)(z_2-z_3))^m$: on the sphere, Haldane recognized this, showing that the 
only states allowed are $|(L^2)L,L,0 \rangle$ where $L=1,3,5,...$ by antisymmetry.  Thus any attempt
to build more general scalar wave functions based on simple products of three-electron factors is bound to fail.

However one can construct a
four-electron state in which this correlation is embedded -- this means that three-electron correlations with
fewer than six quanta, $(m,p_2,p_3)=(1,0,0)$ and $(1,1,0)$ do not appear.  
GH accomplished this as follows,
\begin{eqnarray}
&&\mathcal{S} \left[ (z_1-z_3)(z_1-z_4)(z_2-z_3)(z_2-z_4)  [z_1 z_2]_1 \odot [z_3 z_4]_1 \right]  \nonumber \\
&&= \mathcal{S} \left[ \left\{ (z_1-z_3)^2 (z_1-z_4) \right\} \left\{(z_2-z_3) (z_2-z_4)^2 \right\} \right.+ \nonumber \\
&&~~~~~~\left. +\left\{(z_1-z_3)(z_1-z_4)^2 \right\} \left\{ (z_2-z_3)^2 (z_2-z_4) \right\} \right] \nonumber \\
&&\equiv \mathcal{S} \left[g_6[\{z_1,z_2\},\{z_3,z_4\}] \right] \nonumber  \sim g_6^\prime[z_1,z_2,z_3,z_4]
\label{eq:N4}
\end{eqnarray}
where $[z_1 z_2]_1 \odot [z_3 z_4]_1 = (z_1-z_3)(z_2-z_4) + (z_1-z_4)(z_2-z_3)$, the ``spreading
operator" of GH that plays a role for pairs of electrons analogous to Laughlin's $(z_i-z_j)$ for
individual electrons.  $g_6^\prime$,
before symmetrization, is a simple product of two of the three-electron correlations associated with $f_3^\prime$.
The four-electron correlation is a scalar, and thus the antisymmetric state formed from
it is a scalar: any three electrons selected from this state will have six quanta.  
One can contrast this four-electron plaquette
with its $m=3$ Laughlin counterpart
\begin{equation}
(z_1-z_2)^2 \left[  (z_1-z_3)(z_1-z_4)(z_2-z_3)(z_2-z_4)  \right]^2 (z_3-z_4)^2
\end{equation}
With respect to Laughlin's construction, Eq. (\ref{eq:N4}) replaces
separated pairs of electrons in relative $f$ waves by their $p$ counterparts, and deduced the number of
quanta acting between the pairs from 12 to 10, in the antisymmetric state.  The latter
determines the filling, changing it from 1/3rd- to 2/5ths-filled.  GH, in analogy with Laughlin and Jain,
then wrote this 4-electron wave function as an operator on the half-filled shell.  In their spherical notation
the four-electron operator is
\begin{equation}
d_1 \odot d_2~ d_3 \odot d_4~ [u(1) u(2)]_1 \odot [u(3) u(4)]_1 L[N=4,2].
\label{eq:op}
\end{equation}

These arguments can be made more precise by deriving the most general form of the four-electron 
scalar wave function.  The details will not be given here,
as the analogous construction on a sphere can be found in Ref. \cite{GHsym}.  The result has a familiar form,
paralleling our three-electron result
\begin{eqnarray}
 &&~~~~~~~~~~L[N=4,m] ~ g_4^{\prime~p_4}~ g_6^{\prime~p_6} \nonumber \\
 && ~~ m \in \{ \mathrm{odd~integers} \}~~~~p_4 \in \{0,1,2 \}~~~~p_6 \in \{\mathrm{integers} \} \nonumber \\
 && ~~~~ g_4^\prime = {1 \over 12} g_2^2 -{1 \over 4} g_1 g_3 + g_4 \nonumber \\
&&  ~~~~ g_6^\prime = {2 \over 27} g_2^3 -{1 \over 3} g_1 g_2 g_3 +g_3^2 +g_1^2 g_4 -{8 \over 3} g_2 g_4
   \label{eq:4b}
   \end{eqnarray}
 where $g_1,~g_2,~g_3,~g_4$ are the elementary polynomials for four particles, $g_1=z_1+z_2+z_3+z_4$, ...,
 $g_4= z_1 z_2 z_3 z_4$.   We will ignore for the moment $g_4^\prime$, which we recognize as an $N=4$ 
 scalar in which the $N=3$ correlation $f_2^\prime$ is embedded
 \begin{eqnarray}
     g_4^\prime &\sim& [z_1 z_2 z_3 z_4]^2 \odot [z_1 z_2 z_3 z_4]^2 \nonumber \\
     &\sim& \left[ [z_1 z_2 z_3]^{3/2} \otimes [z_1 z_2 z_3]^{3/2} \right]^1 \odot [z_4]  ^1
 \label{eq:g4}
 \end{eqnarray}
 but which is rather
 uninteresting because it acts as a ``shift" operator.  This leaves $g_6^\prime$ as the interesting new
 correlation, which we already observed is built on the three-electron correlation $f_3^\prime$. 
 More explicitly,
  \begin{eqnarray}
     g_6^\prime &\sim& [z_1 z_4] \odot [z_2 z_3] ~ [z_2 z_4] \odot [z_1 z_3] ~[z_3 z_4] \odot [z_1 z_2] \nonumber \\
     &\sim&  \Big[ [[z_1]\otimes [z_2 z_3]^1]^{1/2} \otimes [[z_2] \otimes [ z_1 z_3]^1]^{1/2} \otimes  \nonumber \\
    &&~~~ [[z_3] \otimes [z_1 z_2]^1]^{1/2} \Big]^{3/2} 
      \odot [z_4]^{3/2}
 \label{eq:g6}
 \end{eqnarray}
 Eqs. (\ref{eq:g4}) and (\ref{eq:g6}) demonstrate that there is a one-to-one correspondence between the translationally
 invariant 3-electron correlations and four-electron scalar wave functions -- the latter is given by the former
 dotted into vectors formed from $z_4$.  This is an important result.  In the simpler case Laughlin investigated, the
 two-electron correlations were themselves scalars, and once one has scalars, it is easy to build many-electron
 wave functions from them.  Here we find that all of
 the physics contained in three-electron correlations maps uniquely into four-electron scalars.  Thus in analogy to 
 Laughlin's construction, many-electron wave functions respecting any three-electron correlation can be constructed
 by using the corresponding four-electron scalar building blocks.
 
 \begin{figure}
\begin{center}
\includegraphics[width=0.48\textwidth]{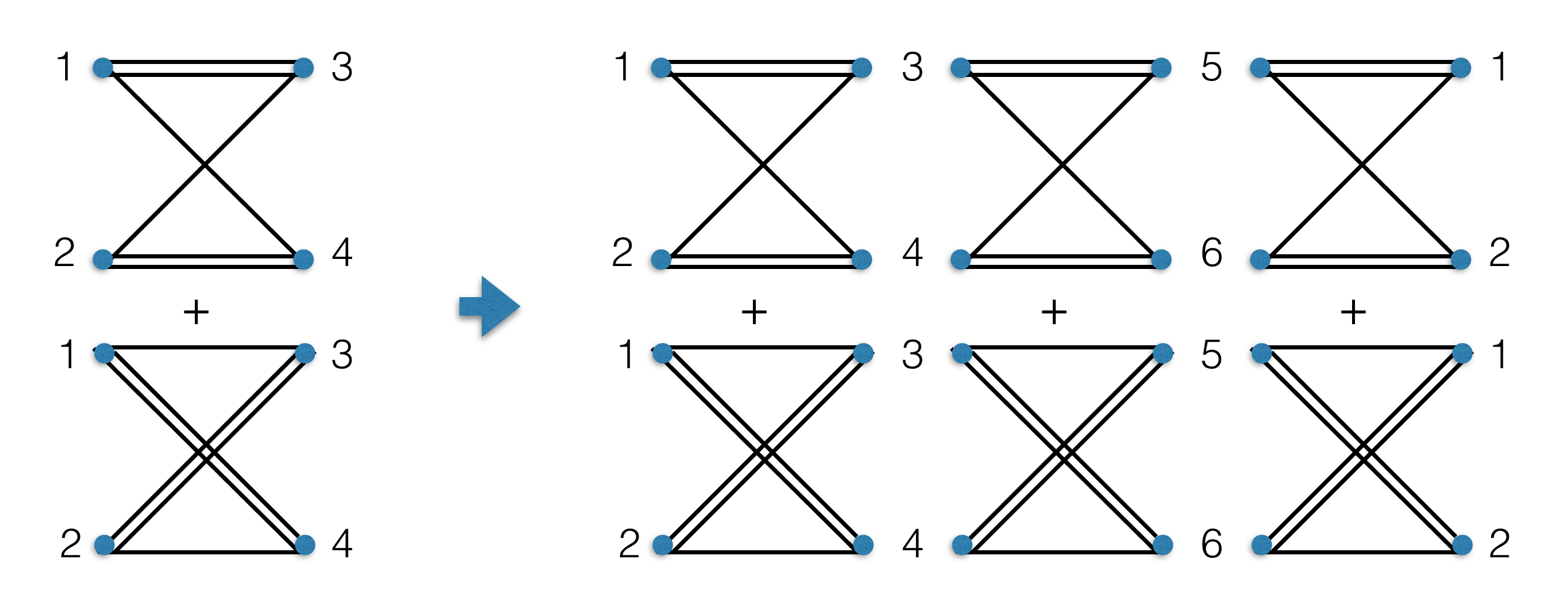}
\end{center}
\caption{The translationally invariant $N=4$ $\nu=2/5$ wave function is unique and has the property
that every three-electron correlation contains six quanta.  This restriction on the short-range behavior
of the wave function can be imposed on configurations of arbitrary $N$: the pattern is shown for $N=6$,
generating a valid scalar wave function when antisymmetrized.  The text notes that wave functions
optimized for their short range properties are much less successful numerically, illustrating that the
scaleless Coulomb potential is not a short-range interaction.}
\label{fig:Fig25Cor}
\end{figure}

This four-electron wave function is unique and thus corresponds to both the GH/Jain and GH$^2$ $N=4$
$\nu=2/5$ results, which are identical.  Because this state is self-conjugate, it also 
corresponds to the $N=4$ $\nu=2/3$ GH and GH$^2$ wave functions.  As this wave function
guarantees that all three-electron correlations contain at least six quanta, it is then tempting to try to
extend the construction to any even $N$, preserving this property in the general $\nu=2/5$ wave
function.  This would then yield an approximate $\nu=2/5$ FQHE wave function that is the exact
eigenfunction of a short-range interaction, effectively generalizing the Haldane potential.  (Clearly
by combining delta functions with gradients, a three-body potential that vanishes among three electrons containing
six quanta, but not among electrons correlated at even shorter distances, can be constructed.)

This in fact can be done, using the four-electron plaquettes described previously.  As the construction is 
much easier to describe geometrically than algebraically, it is presented in Fig. \ref{fig:Fig25Cor}.
This N-electron 2/5-filled state follows the 4-electron form of Eq. (\ref{eq:N4})
\begin{eqnarray}
\Psi^\mathrm{cor}_{2/5}& \sim& L[N,1] ~ \mathcal{S} \left[ \prod_{i<j=1}^{N/2} g_6[I_i,I_j] \right] \nonumber \\
&\equiv& L[N,1] \mathcal{S} \left[ G_6[I_1, \dots I_{N/2}] \right]
\end{eqnarray}
where the $N$ electrons have been partitioned into $N/2$ pairs labeled by $\{I_i\}=\{ \{1,2\},\{3,4\},...,\{N-1,N\} \}$.
The wave function is symmetric by construction under the interchanges $1 \leftrightarrow 2,~3 \leftrightarrow 4,~$etc.,
so it is sufficient to evaluate $\mathcal{S}$ by symmetrizing the wave function over all 
electron interchanges that create distinct partitions of the $N/2$ pairs.
In addition to being the eigenstate of a short-range effective Hamiltonian, this wave function also
has a simple recursion relation.

Despite its nice properties, such a wave function, optimized for its short-range behavior, is not an obvious candidate
for a LO wave function.  As the Coulomb potential has no scale, it would be surprising if its physics could be captured in a short-range
interaction.   The GH$^2$ and GH or Jain $\nu=2/5$ wave functions generate nonzero correlations
among three electrons carrying five quanta, for $N>4$.  We verified that these wave functions, for $N=6$,
have significantly better overlaps with the exact ground state wave function, than does the one we constructed
via the pattern illustrated in Fig. \ref{fig:Fig25Cor}.   This test, it seems to us, is definitive, as there is no arbitrariness in the short-range
wave function: the four-body correlation one must use in a $\nu=2/5$ construction to eliminate configurations with
fewer than six quanta is unique.  

It remains true that wave functions like those of GH/Jain and GH$^2$, that are not optimized for their short-range
behavior, nevertheless can constrain short-range correlations.  Laughlin's wave function is an example, as one has the 
Haldane pseudopotential.  Similarly, the GH/Jain and GH$^2$ wave functions contain no three-body correlations
containing fewer than five quanta, though obviously they are not uniquely defined by this condition.

Thus the example worked through here supports the naive observation that because the Coulomb potential lacks a
short-range scale, good approximate wave functions should not be optimized solely on considerations connected with
short-range interactions.

\end{document}